\documentclass[nofootinbib,prb,twocolumn,amsmath,amssymb,aps]{revtex4-1}
\usepackage{graphicx}
\usepackage{hyperref}
\usepackage{comment}
\usepackage{subcaption}
\usepackage{float}
\hypersetup{
     colorlinks   = true,
     linkcolor    = blue,
     citecolor    = blue
}
\usepackage[all]{hypcap}
\begin{document}

\title{Anomalies and unnatural stability of multi-component Luttinger liquids \\ in $\mathbb{Z}_n\times\mathbb{Z}_n$ spin chains }
\author{Yahya Alavirad}
\author{Maissam Barkeshli}
\affiliation{Department of Physics, Condensed Matter Theory Center and the Joint Quantum Institute, University of Maryland, College Park, MD 20742, USA}

\date{\today}

\begin{abstract}
  We study translationally invariant spin chains where each unit cell contains an $n$-state projective representation of a $\mathbb{Z}_n\times\mathbb{Z}_n$ internal symmetry, generalizing the spin-1/2 XYZ chain. Such spin chains possess a generalized Lieb-Schulz-Mattis (LSM) constraint, and we demonstrate that certain $(n-1)$-component Luttinger liquids possess the correct anomalies to satisfy these LSM constraints. For $n = 3$, using both numerical and analytical approaches, we find that such spin chains with nearest neighbor interactions appear to be gapless for a wide range of microscopic parameters and described by a two-component conformally invariant Luttinger liquid. This implies the emergence of $n-1$ conserved $U(1)$ charges from only discrete microscopic symmetries. Remarkably, the system remains gapless for an unnaturally large parameter regime despite the apparent existence of symmetry-allowed relevant operators in the field theory. This suggests that either these spin chains have hidden conserved quantities not previously identified, or the parameters of the field theory are simply unnatural due to frustration effects of the lattice Hamiltonian. We argue that similar features are expected to occur in: (1) $\mathbb{Z}_n\times\mathbb{Z}_n$ symmetric chains for $n$ odd, and (2) $\mathbb{S}_n\times\mathbb{Z}_n$ symmetric chains for all $n > 2$. Finally, we suggest the possibility of a lower bound growing with $n$ on the minimum central charge of field theories that possess such LSM anomalies. 
\end{abstract}
\maketitle

\section{Introduction}

The Lieb-Schulz-Mattis (LSM) theorem and its generalizations\cite{l1,l2,l4,l5, wen1,tasaki,parameswaran2013,r6,r8,r13,r16,r18,bb1} provide strong constraints on the possible low energy, long wavelength behavior of translationally invariant many-body systems. These theorems can be used, for example, to rule out trivial gapped many-body states of matter in systems with a spin-1/2 degree of freedom per unit cell. Recently, it has been understood that the LSM theorem can be significantly strengthened\cite{bb1} by viewing the system of interest as the boundary of a crystalline symmetry-protected topological (SPT) state \cite{spt1,spt2,spt3}. The bulk-boundary correspondence can then be used to place strong constraints on the type of long wavelength, universal behavior exhibited by the system.

In the framework of quantum field theory, this bulk-boundary correspondence is an example of a 't Hooft anomaly. That is, systems with LSM constraints possess a certain type of 't Hooft anomaly -- a mixed anomaly between the on-site symmetry and the translational symmetry -- and therefore any possible field theory that emerges at long wavelengths must match this anomaly. For example, Refs.~\onlinecite{oshikawa1,oshikawa2} used the relation between LSM constraints and 't Hooft anomaly matching to constrain the possible renormalization group flows that describe critical phases of $SU(N)$ spin chains. Ref.~\onlinecite{bb1} showed how these ideas can be used to constrain the possible patterns of symmetry fractionalization that can arise in (2+1) dimensional topologically ordered states. See also Ref. \onlinecite{komargodski2019,komargodski2018,metlitski} for a discussion of anomalies in field theories of certain quantum magnets. 

In general, given a quantum many-body system that is subject to a generalized LSM constraint, little is known about the space of possible effective field theories that have the appropriate anomalies to satisfy the LSM constraints. To study this question, in this paper we study translationally invariant one-dimensional spin chains with a $\mathbb{Z}_n\times\mathbb{Z}_n$ on-site (internal) symmetry, where each unit cell contains an $n$-dimensional projective representation of  $\mathbb{Z}_n\times\mathbb{Z}_n$. These spin chains can be thought of as generalizations of the XYZ spin chain to $SU(n)$ spin systems. Generalized LSM theorems forbid these spin chains from having a unique gapped ground state\cite{wen1,tasaki} and thus raise an interesting question of what the possible long wavelength effective field theories can be. 

Here we demonstrate that certain $(n-1)$-component Luttinger liquids can satisfy the LSM constraints and thus possess the appropriate 't Hooft anomaly. Importantly, these gapless theories are distinct from the more familiar coupled Luttinger liquids as they have different anomalies and symmetry actions. Moreover, in the case $n = 3$, we use a combination of numerical and analytical methods to study such spin chains in detail. For nearest neighbor models with time-reversal and inversion symmetry, we provide evidence that these systems are indeed described by two-component Luttinger liquids in a large portion of their phase diagram. These critical field theories can be thought of as symmetry-preserving marginal deformations of $SU(3)_1$ Wess-Zumino-Witten (WZW) conformal field theories (CFTs) that are parametrized by a single continuous parameter $g$. Remarkably, these systems possess only discrete internal symmetry generators, and yet have two emergent $U(1)$ charges.

The stability of these gapless phases is particularly surprising. On the one hand, the two-component Luttinger liquids that we study apparently always have symmetry-allowed relevant operators away from the $SU(3)$ invariant point, and therefore strictly speaking are not fully stable as field theories. Nevertheless, remarkably we find that these critical phases of the spin chain are, to within numerical accuracy, stable with respect to tuning the microscopic parameters over a large region of the phase diagram. Specifically, the microscopic parameters can vary over an essentially infinite range, in dimensionless units, with the gapless behavior persisting throughout. In this sense it appears that the parameters of the field theory are ``unnatural'' given the symmetries of the Hamiltonian. We suggest that this unexpected behavior is related to the frustrated nature of terms in the microscopic Hamiltonian, which individually possess a $U(2)^\infty$ symmetry in the thermodynamic limit. Our results suggest that either (1) these spin chains have previously unnoticed conserved quantities -- or may even be fully integrable -- that rule out the appearance of the relevant operators at long wavelengths or (2) the field theory parameters are simply unnatural due to the frustration of the microscopic terms in the Hamiltonian.

We further argue that similar phenomena are expected to occur in: (1) $\mathbb{Z}_n\times\mathbb{Z}_n$ symmetric chains for $n$ odd, and (2) $\mathbb{S}_n\times\mathbb{Z}_n$ symmetric chains for all $n > 2$.

We note that extended gapless phases in the vicinity of the $PSU(3)$ and $PSU(4)$ symmetric spin chains have been previously reported in the literature\cite{extended1,extended2}. However, these models are different from the systems discussed in this work as they do not have the $\mathbb{Z}_n \times \mathbb{Z}_n$ on-site symmetry, but rather have $PSU(2)$ symmetries that prohibit the relevant operators from appearing at long wavelengths and thus stabilize the gapless phases.\footnote{Note that the symmetries involved are $PSU(n) = SU(n)/\mathbb{Z}_n$, as the center $\mathbb{Z}_n$ subgoup of $SU(n)$ acts trivially on all operators.}

The rest of this paper is organized as follows. In Sec.~\ref{s1}, we set up notation and formalism and briefly discuss the anomaly associated with the generalized LSM constraints. In Sec.~\ref{s2}, we review $\mathbb{Z}_2\times\mathbb{Z}_2$ symmetric spin chains and their gapless phases. In Sec.~\ref{s3} we discuss the $\mathbb{Z}_3 \times \mathbb{Z}_3$ spin chains, present numerical evidence for it being gapless over a large region of the phase diagram, and present the two-component Luttinger liquid theory along with its symmetry actions and anomalies. In Sec. \ref{stabilitySec} we discuss in detail the stability of the gapless phase, presenting additional evidence that the two-component Luttinger liquid theory is the correct description despite the presence of symmetry-allowed relevant operators. In Sec.~\ref{ZnSec} we generalize the discussion to $\mathbb{Z}_n\times\mathbb{Z}_n$ symmetric spin chains. We end with a brief discussion in Sec.~\ref{s5}.

\section{Hilbert space, global symmetry operators and the LSM anomalies}
\label{s1}

For each site $i$, the local Hilbert space is an $n$ dimensional space spanned by $|g_i \rangle$, with $g_i = 0, \cdots, n-1$.
We work with onsite ``clock'' and ``shift" operators, $Z_i$ and $X_i$, which form a projective representation of  $\mathbb{Z}_n\times\mathbb{Z}_n$,
\begin{align}
  Z_i|g_i\rangle &= e^{2\pi i g_i/n} |g_i\rangle
  \nonumber \\
  X_i|g_i\rangle &= |(g_{i} + 1) \text{ mod n}\rangle, 
\end{align}
such that:
\begin{align} \label{cmt}
  X_i X_j &= X_j X_i
    \nonumber \\
  Z_i Z_j &= Z_j Z_i
  \nonumber \\
  Z_i X_j &= e^{i2\pi \delta_{i,j}/n} X_j Z_i.
\end{align}
For $n=2$ these are the usual $SU(2)$ Pauli $X_i, Z_i$ operators, corresponding to $\pi$ rotations around the $x$ and $z$ directions.

For a spin chain with $N$ sites, we take the generators of the global $\mathbb{Z}_n\times\mathbb{Z}_n$ symmetry to be given by
\begin{align}
Z=\prod_{i=1}^N Z_i \quad \text{and} \quad X=\prod_{i=1}^N X_i.
\end{align}
Since each site forms a projective representation of $\mathbb{Z}_n \times \mathbb{Z}_n$, a translationally
invariant spin chain that respects these symmetries cannot have a trivial gap (generalized LSM restriction)\cite{wen1,tasaki}.
The ground state must either (1) exhibit spontaneous symmetry breaking of the $\mathbb{Z}_n \times \mathbb{Z}_n$ and/or translational
symmetry, or (2) form a gapless phase. 

One way to see the existence of an anomaly associated with an LSM constraint is as follows. Consider a $\mathbb{Z}_n\times\mathbb{Z}_n$
and translationally invariant spin chain with a unique ground state. From the $\mathbb{Z}_n\times\mathbb{Z}_n$ symmetry we know that
this state has to be a $\mathbb{Z}_n\times\mathbb{Z}_n$ singlet (since it is the unique ground state). However, a singlet state only exists if
the number of sites $N$ is a multiple of $n$, i.e. $N ~\text{mod}~ n =0$. When $N \text{ mod } n = 1$, for instance, the ground state
degeneracy must be a multiple of $n$, and each state must transform non-trivially under the action of $\mathbb{Z}_n \times \mathbb{Z}_n$. 

The above property must be reproduced within the low energy effective field theory. A translation by one lattice site is represented in the field
theory by a non-trivial symmetry transformation on the fields. Changing the number of sites in a periodic system is then modeled by inserting a unit
of flux associated with this translational symmetry, which corresponds to twisting the boundary conditions on the fields. In the presence of this
symmetry flux, the vacuum degeneracy must change, with the $\mathbb{Z}_n \times \mathbb{Z}_n$ symmetry acting non-trivially on the resulting
vacuum subspace. In field theory terminology, this means that the symmetry is broken in the presence of background fields, which in this case corresponds
to the presence of non-trivial translational symmetry flux. This is the symptom of an anomaly, and it implies that the low energy effective field theory cannot
be trivial, thus ruling out a trivial gapped state. More precisely, the field theory must have a mixed anomaly between the $\mathbb{Z}_n \times \mathbb{Z}_n$
internal symmetry and the translational symmetry, which we will refer to below as the LSM anomaly. We will use this particular manifestation of the anomaly
extensively throughout the rest of this work.

Alternatively, we can consider this system to exist at the surface of a (2+1) dimensional weak SPT, corresponding to an array of
(1+1)-dimensional SPTs with $\mathbb{Z}_n \times \mathbb{Z}_n$ symmetry.\cite{bb1} Each (1+1)D SPT in the array is a gapped system with
a linear representation of $\mathbb{Z}_n \times \mathbb{Z}_n$ per unit cell, with boundary zero modes forming a projective representation
of $\mathbb{Z}_n \times \mathbb{Z}_n$. Our (1+1)D spin chain can thus be thought of as a one-dimensional array of these projective
boundary zero modes. As such, the LSM constraint can be interpreted as the impossibility of trivially gapping out the boundary of a (2+1)D weak SPT.

\section{Nearest neighbor Hamiltonian with $\mathbb{Z}_2\times\mathbb{Z}_2$ symmetry}
\label{s2}

In this section we discuss nearest neighbor spin chains with $\mathbb{Z}_2\times\mathbb{Z}_2$ symmetry. While none of the results of this section are new, our method of observing the anomaly is different from standard treatments and sets the stage for the generalization to $\mathbb{Z}_n \times \mathbb{Z}_n$. 

\subsection{Lattice model and symmetries}

The most general nearest neighbor  Hamiltonian with $\mathbb{Z}_2\times\mathbb{Z}_2$ symmetry is given by
\begin{align}
  \label{n2Model}
H=\sum_i \Big( J_x X_i X_{i+1} +  J_y Y_i Y_{i+1} + J_z Z_i Z_{i+1} \Big),
\end{align}
where $Y_i=i X_i Z_i$, and $X$, $Y$, $Z$ are the usual Pauli matrices. This model is known to be integrable for all values of $J_x$, $J_y$, $J_z$. 
In addition to the translational symmetry group $\mathbb{Z}_{\text{trans}}$, generated by translation by one site, $T_x$, and the
$\mathbb{Z}_2\times\mathbb{Z}_2$ symmetry, this system has two important additional discrete symmetries:

\begin{itemize}
\item Inversion, $\mathbb{Z}_2^{P}$, generated by $P$:
\begin{align}
P: ~(X_i,Y_i,Z_i)\rightarrow (X_{-i},Y_{-i},Z_{-i})
\end{align}
\item  Time-reversal, $\mathbb{Z}_2^{\Theta}$, generated by complex conjugation in the $Z_i$ basis, $\Theta$:
\begin{align}
\Theta: ~(X_i,Y_i,Z_i)\rightarrow (X_{i},-Y_{i},Z_{i}).
\end{align}
\end{itemize}
Time-reversal defined here is different from the conventionally defined time-reversal by a unitary $\pi$ rotation around
the $y$ axis. We work with this definition as it can be easily generalized to the $\mathbb{Z}_n\times\mathbb{Z}_n$ case.
Therefore the total symmetry group is
$[\mathbb{Z}_2\times\mathbb{Z}_2 \times \mathbb{Z}_2^{\Theta}] \times [\mathbb{Z}_{\text{trans}}  \rtimes \mathbb{Z}_2^{P}]$. Note
that while the on-site symmetry group $\mathbb{Z}_2\times\mathbb{Z}_2 \times \mathbb{Z}_2^{\Theta}$ is Abelian, the local Hilbert space
on each site forms a projective representation of this symmetry group, so that the representations of the symmetry generators no longer commute. 

For certain special choices of parameters, the model has enhanced symmetries. For example, when $|J_x| = |J_y|$, the $\mathbb{Z}_2 \times \mathbb{Z}_2$
is enhanced to a $U(1) \rtimes \mathbb{Z}_2$ unitary on-site symmetry. When $J_x = J_y = J_z$, the $\mathbb{Z}_2 \times \mathbb{Z}_2$ is enhanced
to the full $SO(3)$ spin rotational symmetry. 

The Hamiltonian in Eq. \ref{n2Model} is exactly solvable and is known to be gapless only when the microscopic on-site symmetry of the system
possesses a continuous $U(1)$ symmetry such that two of the coupling constants are equal in magnitude,
e.g. $J_x=\pm J_y$ (this is a necessary but not sufficient condition for gaplessness). Otherwise the system exhibits spontaneous symmetry-breaking \cite{giamarchi}.

\subsection{Luttinger liquid theory}

The critical phase here is described by the usual $c=1$ compactified boson (Luttinger liquid) CFT,
\begin{align}
L=\frac{1}{2\pi}\int_0^{2\pi} dx \big[ &(\partial_t \varphi(x,t))^2 - (\partial_x \varphi(x,t))^2 \big] ,\\ \newline \nonumber
& \varphi \sim \varphi+2\pi R
 \end{align}
 To simplify the formalism we have set the length of the chain to $2\pi$. $R$ is compactification radius of the free boson.
 The Luttinger parameter $K$ is related to $R$ as $K^{-1}=4R^2$. The compactified field $\varphi$ can be expanded as:
\begin{align}\label{mde}
&\varphi(x,t)=\varphi_0+ \frac{n}{2R} t +   x m R + \sum_{l\neq 0} \varphi_l(t) e^{ilx},\\ \newline \nonumber
&n,m,l\in \mathbb{Z}.
\end{align}
Ignoring the trivial harmonic oscillator part (setting $\varphi_l=0$), the zero mode Hamiltonian reads,
\begin{align}\label{h0}
H_0=\frac{1}{2} (\frac{n^2 }{2 R^2}+2 m^2 R^2).
\end{align}
 This part of the spectrum sets the scaling dimension of the primary fields. It is useful to decompose the field $\varphi$ into right/left moving parts,
\begin{align}
\varphi_{L/R} (x\pm t) = \varphi_{0,L/R}+ \frac{1}{2}(m R\pm \frac{n}{2R})(x\pm t)  + \text{oscillators}.
\end{align} 
The conventional ($2\pi$ periodic) Luttinger variables are given by,
\begin{align}
&\phi(x)=\frac{1}{R}(\varphi_L(x)+\varphi_R(x)) \\ \newline \nonumber
&\theta(x)=2R (\varphi_L(x)-\varphi_R(x)).
\end{align}

Note that the T-duality $R\rightarrow \frac{1}{2R}$ is manifest in Eq.\eqref{h0}. At the $R=\frac{1}{\sqrt{2}}$ self dual point, the symmetry of the system is enhanced from $U(1) \rtimes \mathbb{Z}_2$ to $PSU(2) \equiv SU(2)/\mathbb{Z}_2 = SO(3)$ and the system is described by the $SU(2)_1$ WZW CFT\cite{yellow}. 

The action of symmetry operators can be read off using the usual bosonization methods\cite{giamarchi,metlitski}.
The $\mathbb{Z}_2$ symmetry associated with $X$ is given by,
\begin{align}
X: ~\theta\rightarrow-\theta ~~\text{and}~~ \phi\rightarrow-\phi,
\end{align}
and the $\mathbb{Z}_2$ symmetry associated with $Z$ is,
\begin{align}\label{zac}
Z:~~\theta\rightarrow\theta+\pi~~\text{and}~~ \phi\rightarrow\phi.
\end{align}
The translational symmetry $T_x$ acts like,
\begin{align}\label{tr1}
T_x:~~\theta\rightarrow\theta+\pi ~~\text{and}~~ \phi\rightarrow\phi+\pi.
\end{align}
Finally the symmetry actions associated with time reversal ($\varphi_L\rightarrow \varphi_R$) and inversion ($\varphi_L\rightarrow -\varphi_R$) can be represented as,
\begin{align}
P:~~\theta \rightarrow \theta  ~~~\text{and}~~ ~~\phi\rightarrow -\phi.
\end{align}
and,
\begin{align}
\Theta:~~\theta \rightarrow -\theta ~~~\text{and}~~ ~~\phi\rightarrow \phi.
\end{align}

\subsection{LSM anomaly}

 Inserting a unit of translational symmetry flux changes the boundary condition (Eq.\eqref{tr1}),
\begin{align}
&\phi(x+2\pi)=\phi(x)+2\pi m +\pi \\ \newline \nonumber
&\theta(x+2\pi)=\theta(x)+2\pi n +\pi.
\end{align}
This can then be incorporated in the mode expansion Eq.\eqref{mde} as,
\begin{align}\label{trnl}
m\rightarrow m+\frac{1}{2}\quad;\quad n\rightarrow n+\frac{1}{2}.
\end{align}
An implication of this shift is that (after flux insertion) the total spin operator is given by $S_z^{tot}=\frac{1}{2\pi }\int_0^{2\pi} dx \partial_x \phi =m+1/2$. That is, the total $S_z$ charge of the spin chain becomes half-integer, which of course matches the microscopic expectation (spin chains of odd length have half-integer $S_z^{tot}$). 

 Eq.\eqref{trnl} shows that for odd length spin chains the ground state is four fold degenerate corresponding to $\{n=\pm1/2,m=\pm1/2\}$. Note that microscopically only a two fold degeneracy is guaranteed and that this four fold degeneracy is a consequence of the fact that the $U(1)$ symmetry is enlarged to $U(1)_L \times U(1)_R$. In practice this degeneracy is always broken to two two-fold degeneracies by irrelevant perturbations that break the symmetry down to $U(1)$. This system size (or translation flux) dependent change in the degeneracy is a manifestation of the mixed $\mathbb{Z}_n\times\mathbb{Z}_n$ and $\mathbb{Z}_{\text{trans}}$ LSM anomaly in the critical phase.

\subsection{Stability of gapless phase}
 
As mentioned above, the LSM theorem forbids a trivial gap (without spontaneous symmetry breaking). However, it does not guarantee the presence of a critical phase or its stability. To analyze stability of the gapless phase, we consider the most generic perturbations consistent with the symmetries that can be added to the critical action described above, and which could potentially gap the system. These are of the form
\begin{align}
\cos (m\theta+n\phi) .
\end{align}
Note that sinusoidal terms are not considered as they are manifestly not invariant under the symmetry actions. The scaling dimension of the operator above is given by Eq.\eqref{h0}. The most relevant operators that are consistent with the anomaly generating symmetry group $\mathbb{Z}_n\times\mathbb{Z}_n\times \mathbb{Z}_{\text{trans}}$ are
\begin{align}
\cos (2\theta) ~~\text{and}~~\cos (2\phi).
\end{align}
These operators are also invariant with respect to $P$ and $\Theta$ symmetries. At the self-dual point $R=\frac{1}{\sqrt{2}}$ both of the operators above are marginal. Therefore, at the self-dual $PSU(2)$ symmetric point the theory is gapless. As we move away from the self dual point one of these operators becomes relevant and pins its argument at strong coupling, giving rise to a gapped phase with $\langle \phi \rangle \neq 0$ or $\langle \theta \rangle\neq 0$. Therefore, away from the self-dual point, the critical phase is not stable. What is needed is a symmetry that prohibits at least one of the two operators above. In the case of spin chains discussed here, such symmetry is present at $U(1)$ symmetric points where two of the coupling constants are equal in magnitude, e.g. $J_x=\pm J_y$. This $U(1)$ symmetry (corresponding to continuous rotations around the $z$ axis) acts as,
\begin{align}
U(1):~~\theta\rightarrow\theta+\alpha ~~\text{and}~~ \phi\rightarrow\phi,
\end{align}
where $\alpha$ is an arbitrary real constant. This symmetry prohibits the $\cos (2\theta)$ term. Therefore, in an extended region $R<\frac{1}{\sqrt{2}}$, no relevant operators are allowed and the critical phase is stable. This of course matches the known result that the spin chains discussed above can be gapless only when an additional $U(1)$ symmetry is present and $R<\frac{1}{\sqrt{2}}$.
 
\section{Nearest neighbor  Hamiltonian with $\mathbb{Z}_3\times\mathbb{Z}_3$ symmetry}\label{s3}

\subsection{Lattice model, symmetries, and phase diagram}

We consider the most general translationally invariant nearest neighbor Hamiltonian with on-site $\mathbb{Z}_3\times\mathbb{Z}_3$ symmetry:
\begin{align}\label{h3}
H=&\sum_i \Big( J_w W_i W^\dagger_{i+1}+ J_x X_i X^\dagger_{i+1} +  J_y Y_i Y^\dagger_{i+1} +J_z Z_i Z^\dagger_{i+1}  \Big)\\ \newline \nonumber
&+h.c~ ,
\end{align}
where $W,Y$ are defined as
\begin{align}
W=Z^\dagger X,\qquad~Y=ZX.
\end{align}
Any pair of these four operators ($W,X,Y,Z$) satisfy relations analogous to Eq.\eqref{cmt}. We can write a matrix representation of these operators (acting on the onsite Hilbert space) as,
\begin{align}\label{mtxel}
&W=\left(\begin{array}{ccc} 0 & e^{4\pi i /3} & 0 \\ 0 & 0 & e^{2\pi i /3} \\ 1 & 0 & 0\end{array}\right) &&X=\left(\begin{array}{ccc} 0 & 1 & 0\\ 0 & 0 & 1 \\ 1 & 0 & 0\end{array}\right)\\ \newline \nonumber 
&Y=\left(\begin{array}{ccc} 0 & e^{2\pi i /3} & 0 \\ 0 & 0 & e^{4\pi i /3} \\ 1 & 0 & 0\end{array}\right) &&Z=\left(\begin{array}{ccc} e^{2\pi i /3} & 0 & 0\\ 0 & e^{4\pi i /3} & 0 \\ 0 & 0 & 1\end{array}\right) .
\end{align}
Note that as opposed to the $\mathbb{Z}_2\times\mathbb{Z}_2$ case, the coupling constants, e.g. $J_z$, do not have to be real.

For various specific choices of coupling constants $J_i$, the model has enhanced symmetries. For example:
\begin{itemize}

\item
If all the coupling constants are \textit{real}, we have an onsite, unitary $\mathbb{Z}^{\mathcal{C}}_2$ charge conjugation symmetry,
\begin{align} 
\mathcal{C}=\left(\begin{array}{ccc} 0 & 1 & 0 \\ 1 & 0 & 0 \\ 0 & 0 & 1\end{array}\right).
\end{align} 
This symmetry acts as
\begin{align}
\mathcal{C}: ~(W_i,X_i,Y_i,Z_i)\rightarrow (e^{2\pi i /3} W^\dagger_i,X^\dagger_{i},e^{-2\pi i /3}  Y^\dagger_{i},Z^\dagger_{i}).
\end{align}
The charge conjugation symmetry enlarges the $\mathbb{Z}_3 \times \mathbb{Z}_3$ on-site symmetry to $[\mathbb{Z}_3 \times \mathbb{Z}_3]\rtimes \mathbb{Z}_2^{\mathcal{C}}$. 
\begin{figure}[t]
\centering
	\vspace{1mm}
\includegraphics[width=0.9\columnwidth,keepaspectratio]{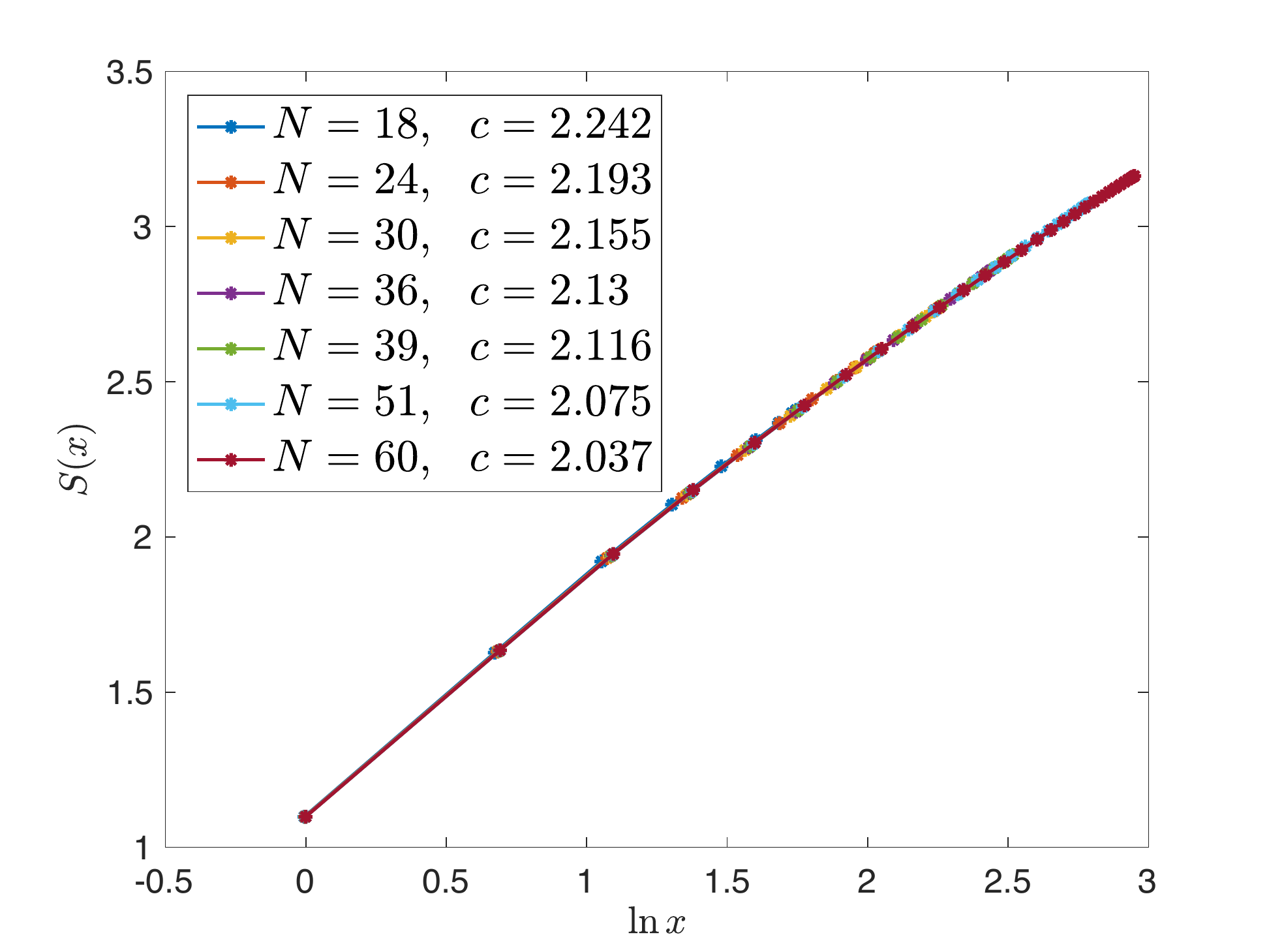} 
\caption{central charge estimate for a sample point ($J_x=J_z=1$ and $J_y=J_w=0$)  in the critical phase of the $\mathbb{Z}_3\times\mathbb{Z}_3$ model. $x$ here is $(\frac{N}{\pi} \sin (\frac{\pi L}{N}))$.}
\label{fig3}
\end{figure}
\item Again, if all the coupling constants are \textit{real}, there is an inversion symmetry, $\mathbb{Z}_2^P$
\begin{align}
P: ~(W_i,X_i,Y_i,Z_i)\rightarrow (W_{-i},X_{-i},Y_{-i},Z_{-i}).
\end{align}

\item
If $J_x$ is real and $J_w=J_y^*$, there is a time-reversal symmetry, $\mathbb{Z}_2^\Theta$, associated with complex conjugation in the $Z_i$ basis,
\begin{align}
\Theta: ~(W_i,X_i,Y_i,Z_i)\rightarrow (Y_i,X_{i},W_{i},Z^\dagger_{i}).
\end{align}
Note that with time-reversal $\mathbb{Z}_2^\Theta$, the $\mathbb{Z}_3 \times \mathbb{Z}_3$ on-site symmetry is expanded to
$\mathbb{Z}_3 \times [\mathbb{Z}_3 \rtimes \mathbb{Z}_2^{\Theta}] = \mathbb{Z}_3 \times \mathbb{S}_3 $. On the other hand,
the $\mathbb{Z}_2$ symmetry generated by $\mathcal{C} \Theta$ by itself also enhances the $\mathbb{Z}_3 \times \mathbb{Z}_3$ on-site symmetry
to $ [\mathbb{Z}_3 \rtimes \mathbb{Z}_2^{\mathcal{C}\Theta}]  \times \mathbb{Z}_3 =\mathbb{S}_3 \times \mathbb{Z}_3 $

\item  If three of the coupling constants are equal, e.g. $J_x=J_y=J_z$, the $\mathbb{Z}_3 \times \mathbb{Z}_3$ on-site symmetry is enhanced to $ \mathbb{Z}_3 \ltimes U(1)^2$. Here the two $U(1)$ symmetries are associated with conservation of the real and imaginary part of $Z$ (i.e. $Z+Z^\dagger$ and $i(Z-Z^\dagger)$).

  \item If all four of the coupling constants are equal, the $\mathbb{Z}_3 \times \mathbb{Z}_3$ on-site symmetry is enhanced to  $PSU(3)=SU(3)/\mathbb{Z}_3$.
  \end{itemize}

The $PSU(3)$ symmetric point has several known integrable deformations\cite{SCHULTZ,VEGA,deform1,deform2,deform3}, however none of these are $\mathbb{Z}_3\times\mathbb{Z}_3$ invariant (and therefore not of the form in Eq.\eqref{h3}), with the exception of the particular deformation shown in Eq. \ref{etaDeformation}.

We further note that Ref.\onlinecite{extended1} has previously reported an extended gapless phase in a particular $PSU(2)$ invariant deformation of the $PSU(3)$ chain. However, Eq.\eqref{h3} is not $PSU(2)$ invariant away from the $PSU(3)$ symmetric point. To see this, we can write Eq.\eqref{h3} in terms of the standard $SU(2)$ spin-1 matrices and confirm that the Hamiltonian is not left invariant under $PSU(2)$ transformations. Therefore, our model is quite different from the one studied in Ref.\onlinecite{extended1}.

We use the DMRG method to study this model numerically\cite{dmrg}. Calculations were performed using the ITensor C++ Library\cite{ITensor}. To estimate the central charge, we usually work with periodic boundary conditions (PBC). While DMRG is notoriously bad at handling PBCs, in practice PBCs can give an accurate estimate of the central charge even for quite small systems. In particular, to detect gapless phases and to calculate their central charge $c$, we fit the numerically calculated entanglement entropy to the CFT form\cite{ee},
\begin{align}\label{ee}
S(L)=\frac{c}{3} \ln (\frac{N}{\pi} \sin (\frac{\pi L}{N}))+a,
\end{align}
where $a$ is a non-universal constant, $N$ is the number of sites and $L$ is the number of sites in the subregion used to calculate the entanglement entropy.

We observe that a significant portion of the phase diagram around the fully symmetric (antiferromagnetic) point $J_x=J_y=J_w=J_z=|J|$ is described by a $c=2$ CFT. A sample plot of how this central charge is extracted is given in Fig.~\ref{fig3}. Due to the large number of parameters in Eq.\eqref{h3}, mapping out the entire phase diagram is impractical. For concreteness, here we discuss this system in a few specific cases. It should be emphasized that the $c=2$ gapless phase is seemingly prevalent all around the fully symmetric anti-ferromagnetic point $J=J_x=J_y=J_w=J_z>0$ and is by no means restricted to the cases considered below.

\subsubsection{Fully symmetric point, $J_x=J_y=J_w=J_z=|J|$, with $PSU(3)$ symmetry}

\begin{figure}[t]
\centering
	\vspace{1mm}
\includegraphics[width=0.92\columnwidth,keepaspectratio]{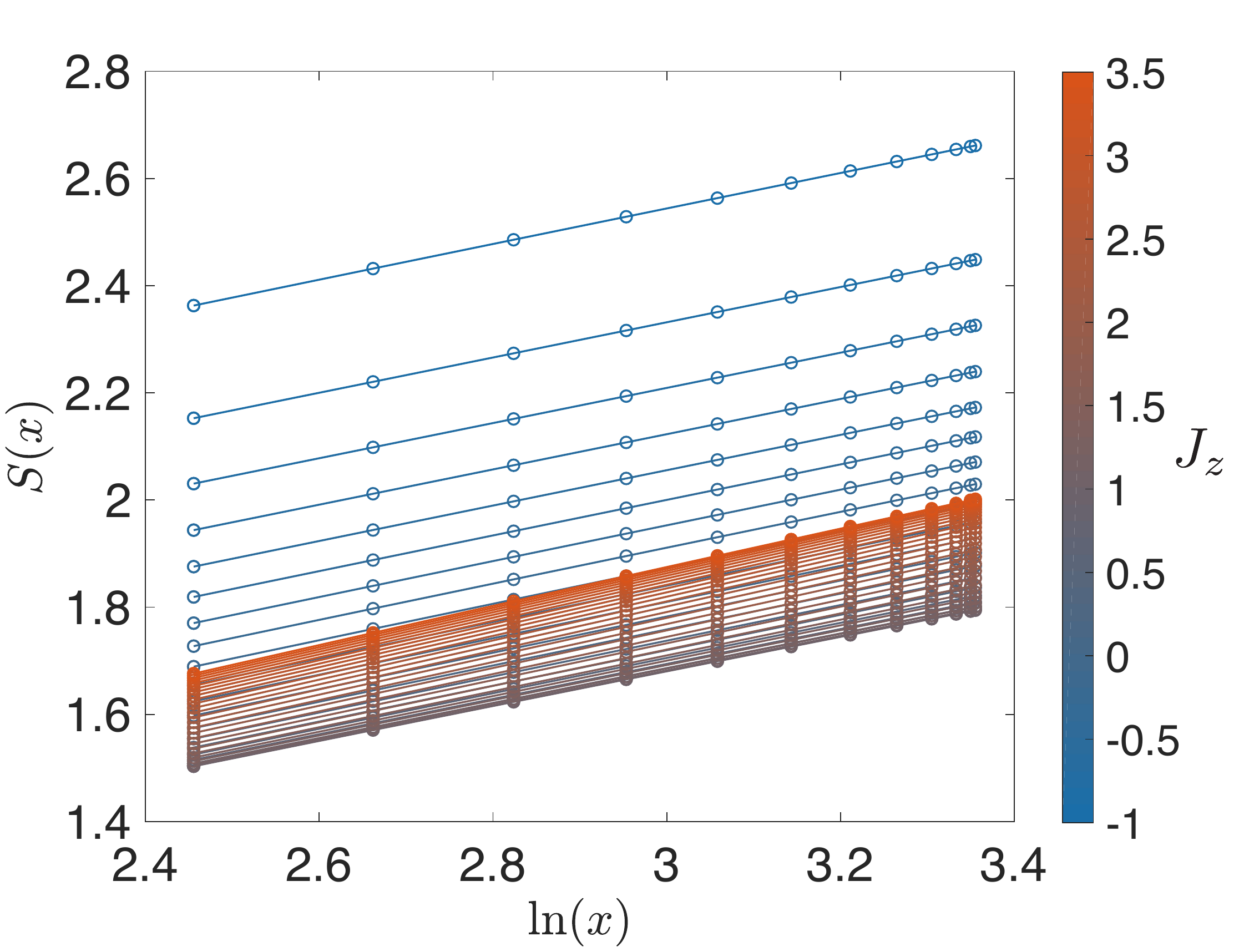} 
\caption{Entanglement entropy of the $U(1)^2$ symmetric model for different values of $J_z$ (value of $J_z$ in color-bar). We have set $J=J_w=J_x=J_y=1$. Here, $x$  is $(\frac{N}{\pi} \sin (\frac{\pi L}{N}))$. All the lines are parallel and therefore, central charge seems constant for $-|J|<J_z$. We have used open boundary conditions at $N=90$.}
\label{eec}
\end{figure}

At the fully symmetric point $J_x=J_y=J_w=J_z=|J|$, the Hamiltonian above is equivalent to the $SU(3)$ Heisenberg spin chain (Lai-Sutherland model\cite{Lai,Sutherland}),
\begin{align}\label{su3}
H=&|J| \sum_i \Big( W_i W^\dagger_{i+1}+ X_i X^\dagger_{i+1} +  Y_i Y^\dagger_{i+1} + Z_i Z^\dagger_{i+1}  \Big)+h.c\\ \newline \nonumber 
=& \frac{3|J|}{2} \sum_i \sum_{j=1}^8 \lambda^j_{i} \lambda^j_{i+1},
\end{align}
where the $\lambda_j$ are the Gell-Mann matrices generating the defining representation of $SU(3)$,
\begin{align}
&\lambda_1=\left(\begin{array}{ccc} 0 & 1 & 0\\ 1 & 0 & 0 \\ 0 & 0 & 0\end{array}\right)~\lambda_2=\left(\begin{array}{ccc} 0 & -i & 0\\ i & 0 & 0 \\ 0 & 0 & 0\end{array}\right)~ \lambda_3=\left(\begin{array}{ccc} 1 & 0 & 0\\ 0 & -1 & 0 \\ 0 & 0 & 0\end{array}\right) \\ \newline \nonumber
&\lambda_4=\left(\begin{array}{ccc} 0 & 0 & 1\\ 0 & 0 & 0 \\ 1 & 0 & 0\end{array}\right)~\lambda_5=\left(\begin{array}{ccc} 0 & 0 & -i\\ 0 & 0 & 0 \\ i & 0 & 0\end{array}\right)~ \lambda_6=\left(\begin{array}{ccc} 0 & 0 & 0\\ 0 & 0 & 1 \\ 0 & 1 & 0\end{array}\right)  \\ \newline \nonumber
&\lambda_7=\left(\begin{array}{ccc} 0 & 0 & 0\\ 0 & 0 & -i \\ 0 & i & 0\end{array}\right)~ \lambda_8=\frac{1}{\sqrt{3}}\left(\begin{array}{ccc} 1 & 0 & 0\\ 0 & 1 & 0 \\ 0 & 0 & -2\end{array}\right) .
\end{align}
This model has two conserved $U(1)$ charges (isospin and hyper-charge) corresponding to  the Cartan sub-algebra of $SU(3)$,
\begin{align}
&S_z=\sum_i -\frac{2i}{\sqrt{3}}(Z_i-Z_i^\dagger)=\sum_i \lambda_i^3 \\ \newline \nonumber
&Q_z=-\frac{1}{2}\sum_i (Z_i+Z_i^\dagger)=\sum_i \frac{\sqrt{3}}{2}\lambda_i^8.
\end{align} 

The $SU(3)$ spin chain is Bethe-ansatz solvable\cite{Sutherland} and known to be described by the $SU(3)_1$ WZW CFT in the long wavelength limit\cite{wzw0,wzw1,wzw2}. Recently, it has been shown that in the continuum description, this spin chain can be described as a sigma model on the flag manifold $SU(3)/U(1)^2$\cite{bykov1,bykov2,lajko}. These works motivated more detailed field theoretical studies of such sigma models\cite{tanizaki1,tanizaki2,seiberg}. Of particular relevance to this paper is the observation of Ref.~\onlinecite{tanizaki1} that the anomalies associated with the $PSU(3)$ symmetric point survive even if the symmetry of the sigma model is broken down to $\mathbb{Z}_3\times \mathbb{Z}_3\subset{PSU(3)}$.
\begin{figure}[t]
\centering
	\vspace{1mm}
\includegraphics[width=0.92\columnwidth,keepaspectratio]{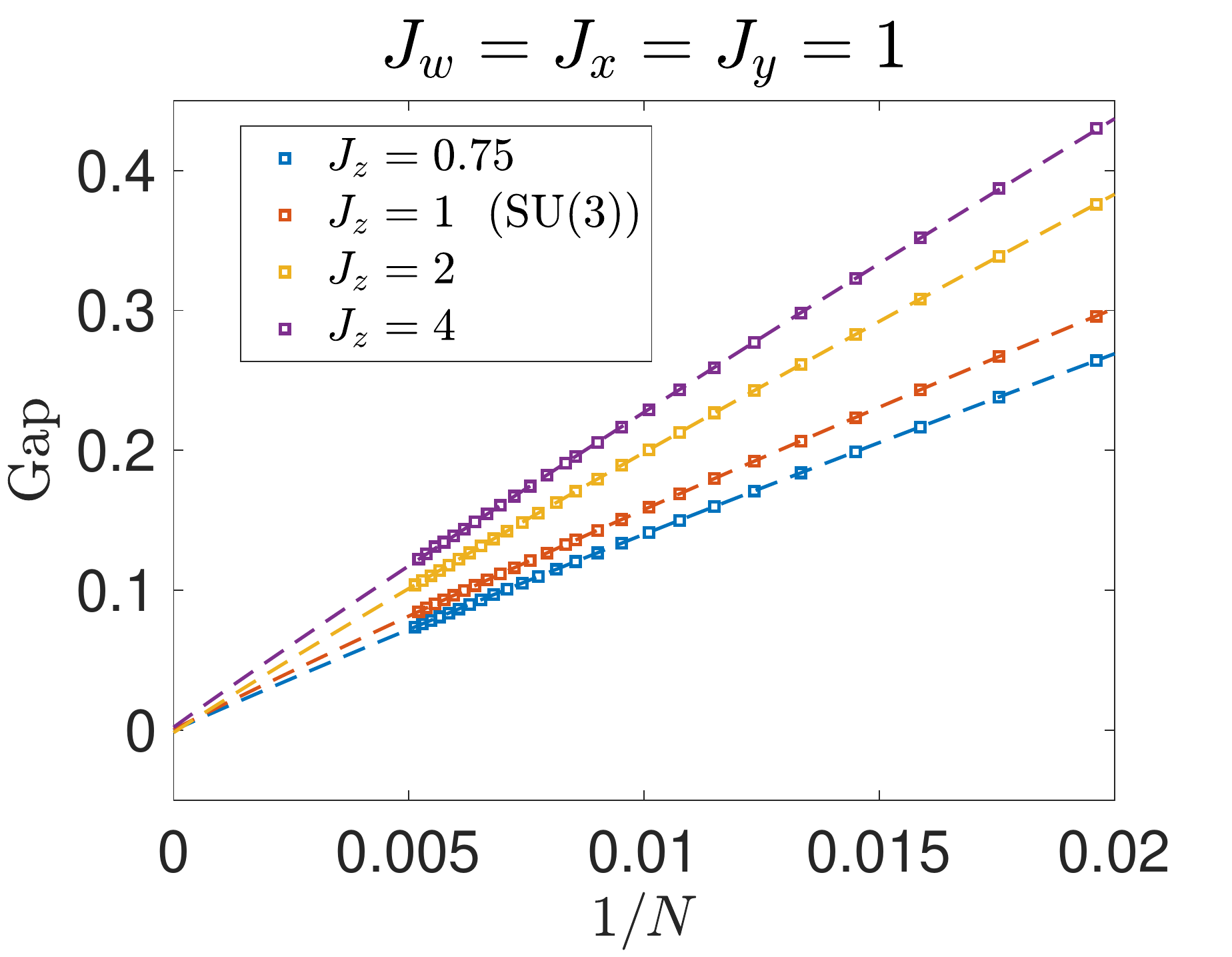} 
\caption{Finite size scaling of the energy difference between $S_z=0$ and $S_z =1$ sectors of the $U(1)^2$ symmetric model for different values of $J_z$ at $J=J_w=J_x=J_y=1$. We use open boundary condition and consider systems sizes up to $N=195$.}
\label{gapnew}
\end{figure}

\subsubsection{$J=J_x=J_y=J_w=\pm 1\neq J_z$ with $U(1)^2$ symmetry}

Another case of interest is when $3$ of the couplings are equal and real, e.g. $J=J_x=J_y=J_w=\pm 1\neq J_z$. In this case the $PSU(3)$ symmetry is broken down to $ \mathbb{Z}_3 \ltimes U(1)^2 $, such that isospin and hyper-charge are still conserved. Based on numerical results, it appears that for $-|J|<J_z$ this model is still described by a $c=2$ CFT (see Fig.\ref{eec}). This case, for $|J|<J_z$, has also been considered in Ref.~\onlinecite{ryu1}. Since the gapless phase here is smoothly connected to the $SU(3)_1$ CFT, it is naturally described by a marginal deformation of it.

Here we also present evidence that the spin chain remains gapless for $J=J_x=J_y=J_w$ and $-|J|<J_z$. We observe that the finite size scaling of the gap is consistent with a vanishing gap in the thermodynamic limit, for both the $J_z > |J|$ and $J_z > - |J|$ regimes. Fig.~\ref{gapnew} shows the energy difference between $S_z=0$ and $S_z =1$ sectors in this regime. We have used a fit of the form $\Delta=\frac{a}{N}+\frac{b}{N\log(N)}+c$, where $a$, $b$, and $c$ are the fitting parameters and the logarithmic contribution arises from the presence of marginal operators in the field theory. 

We have also examined the model for $J_z$ as large as $40 |J|$. Although we have not studied the finite-size scaling of the gap in detail, from the entanglement entropy scaling we note that the model seems to remain gapless. The same conclusion has been reported by Ref. \onlinecite{ryu1} for $J_z$ as large as $100 |J|$. 

\subsubsection{ $J_y=J_w=0$, $\mathbb{Z}_3\times \mathbb{Z}_3$ symmetry }
\begin{figure}[t]
\centering
	\vspace{1mm}
\includegraphics[width=0.6\columnwidth,keepaspectratio]{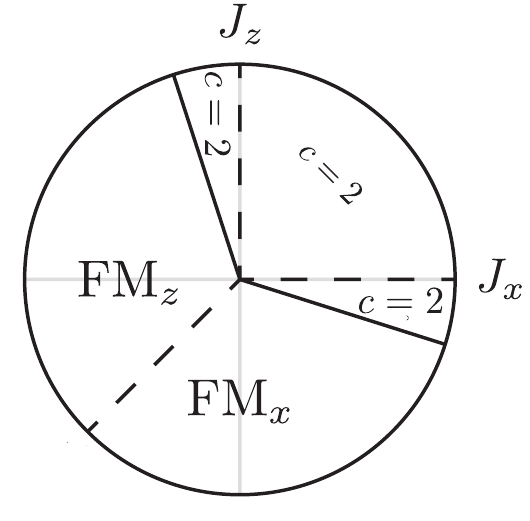} 
\caption{Phase diagram of the $\mathbb{Z}_3\times\mathbb{Z}_3$ model at $J_w=J_y=0$. Dashed lines correspond to first order transitions. Solid lines correspond to second order transitions}
\label{fig1}
\end{figure}

Finally, we consider the case where all the couplings are real and two of them, e.g. $J_y=J_w=0$ are set to zero. In this case the $PSU(3)$ symmetry is broken
all the way down to the discrete group $\mathbb{Z}_3\times \mathbb{Z}_3$. In this case, the Hamiltonian can be written as,
\begin{align}\label{qtc}
H=\sum_i |J|\Big( \sin(\theta) X_i X^\dagger_{i+1} +  \cos(\theta) Z_i Z^\dagger_{i+1}  \Big)+h.c
\end{align}

\begin{figure}[t]
\centering
	\vspace{0.0mm}
\includegraphics[width=\columnwidth,keepaspectratio]{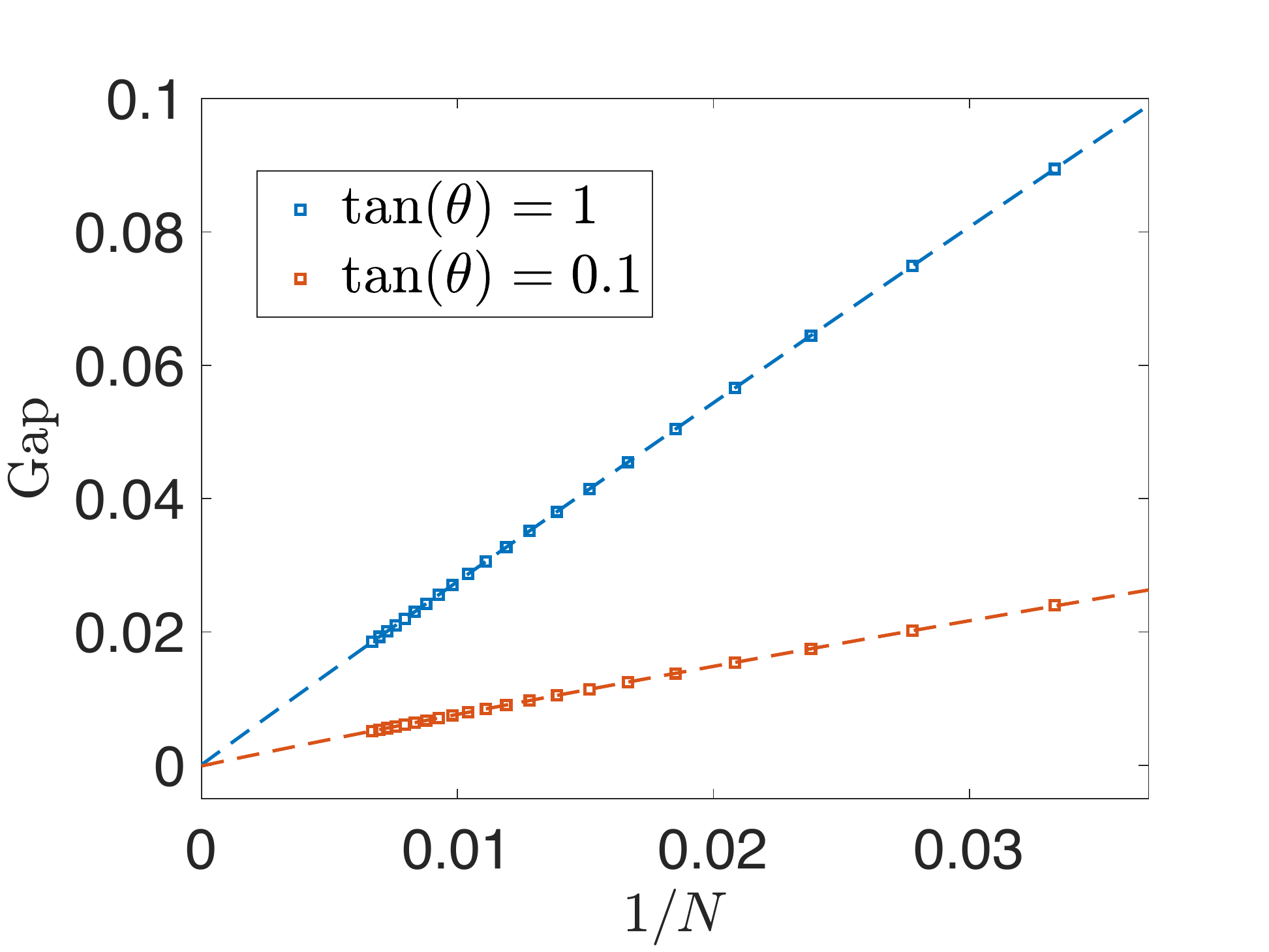} 
\caption{Finite size scaling of the energy difference between $\sum_i g_i ~\text{mod}~ 3 =0$ and $\sum_i g_i~ \text{mod} ~3 =1$ sectors of the quantum torus chain Eq.\eqref{qtc} in the gapless regime. We use open boundary condition and consider systems sizes up to $N=150$. }
\label{fig1prime}
\end{figure}

This Hamiltonian has been previously studied in Ref.~\onlinecite{ryu2} under the title of ``quantum torus chain". The numerically calculated phase diagram is plotted in Fig.~\ref{fig1} (the same phase diagram was found independently in Ref.~\onlinecite{ryu2}).  The two phases $FM_x~\text{and}~ FM_z$ are associated with $\langle X_i \rangle \neq 0$ and $\langle Z_i \rangle \neq 0$, respectively. The central charge is estimated by fitting the entanglement entropy to Eq.\eqref{ee}. As shown in Fig.~\ref{fig1}, this system hosts a large region described by a $c=2$ CFT.  To provide further evidence of being gapless, we have studied the finite size scaling of the energy gap. Fig.~\ref{fig1prime} shows the energy difference between $\sum_i g_i ~\text{mod}~ 3 =0$ and $\sum_i g_i~ \text{mod} ~3 =1$ sectors for two different points  $\theta$ in the gapless phase. As shown in Fig.~\ref{fig1prime} and consistent with the CFT prediction, the energy gaps appear to flow to zero in the thermodynamic limit as  $\Delta=\frac{a}{N}+\frac{b}{N\log(N)}$, where the logarithmic contribution arises from the presence of marginal operators in the field theory.

\begin{figure}[t]
\centering
	\vspace{0.5mm}
\includegraphics[width=0.95\columnwidth,keepaspectratio]{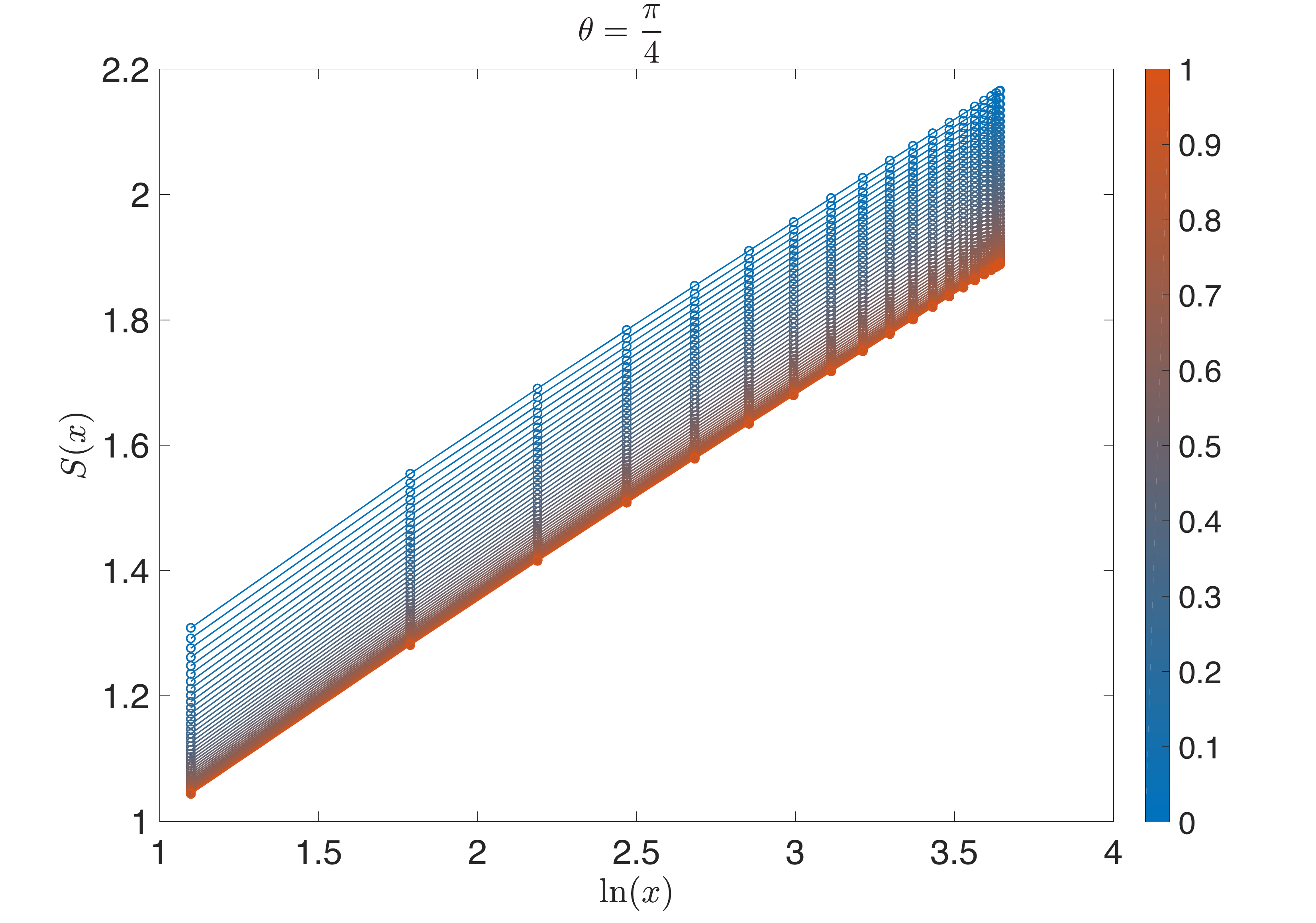} 
\caption{Entanglement entropy for different values of $\lambda$ defined in Eq.\eqref{Hlambda} at $\theta=\pi/4$ (value of $\lambda$ in color-bar). Here, $x$  is $(\frac{N}{\pi} \sin (\frac{\pi L}{N}))$. All the lines are parallel and therefore, central charge is constant as function of $\lambda$. We have used open boundary conditions at $N=120$.}
\label{lambda0.pdf}
\end{figure}

\begin{figure}[t]
\centering
	\vspace{0.5mm}
\includegraphics[width=0.95\columnwidth,keepaspectratio]{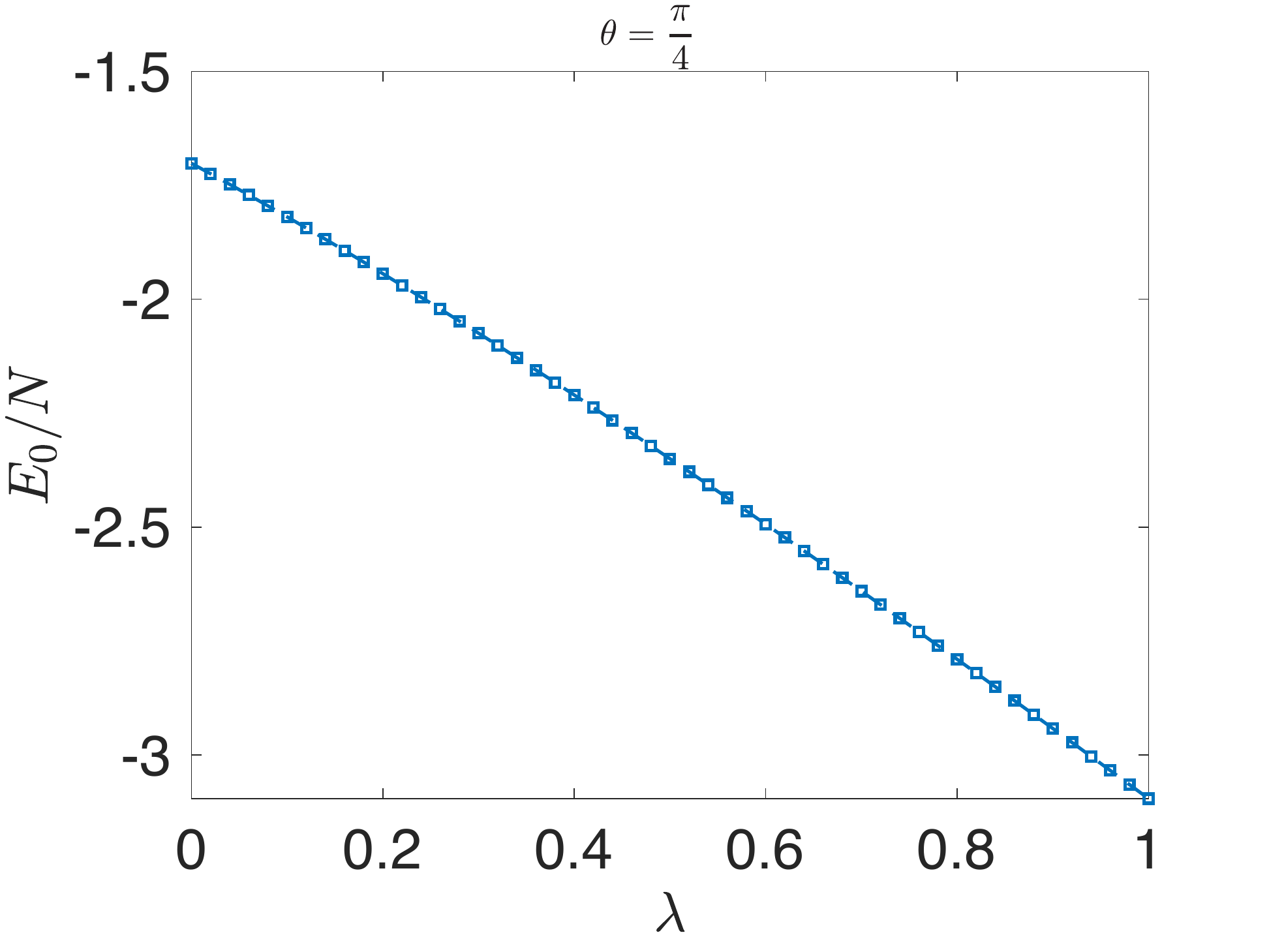} 
\caption{Ground state energy per-site as a function of $\lambda$ defined in Eq.\eqref{Hlambda}  at $\theta=\pi/4$. Here we use open boundary conditions at $N=120$.}
\label{lambda.pdf}
\end{figure}

The nature of the $c=2$ gapless phase was not identified in  Ref.~\onlinecite{ryu2} and left as an open question. Here we demonstrate that this gapless phase can be smoothly deformed into the $PSU(3)$ point without encountering any singularities. Let us consider
the one-parameter family of Hamiltonians obtained by tuning $\lambda$ from $0$ to $1$ in
\begin{align}
  \label{Hlambda}
H_\lambda = \lambda H+(1-\lambda) H_{SU(3)},
\end{align}
with $H_1 = H$ given by Eq. \ref{qtc}. We observe that the central charge as estimated from entanglement entropy remains constant for all values of $\lambda$ (see Fig. \ref{lambda0.pdf}). Furthermore, to ensure no first order transitions occur as we change $\lambda$, we have also calculated the ground state energy (per-site). Sample results are shown in Fig.~\ref{lambda.pdf}, showing no sign of a phase transition. 

\begin{figure}[t]
\centering
	\vspace{0.8mm}
\includegraphics[width=\columnwidth,keepaspectratio]{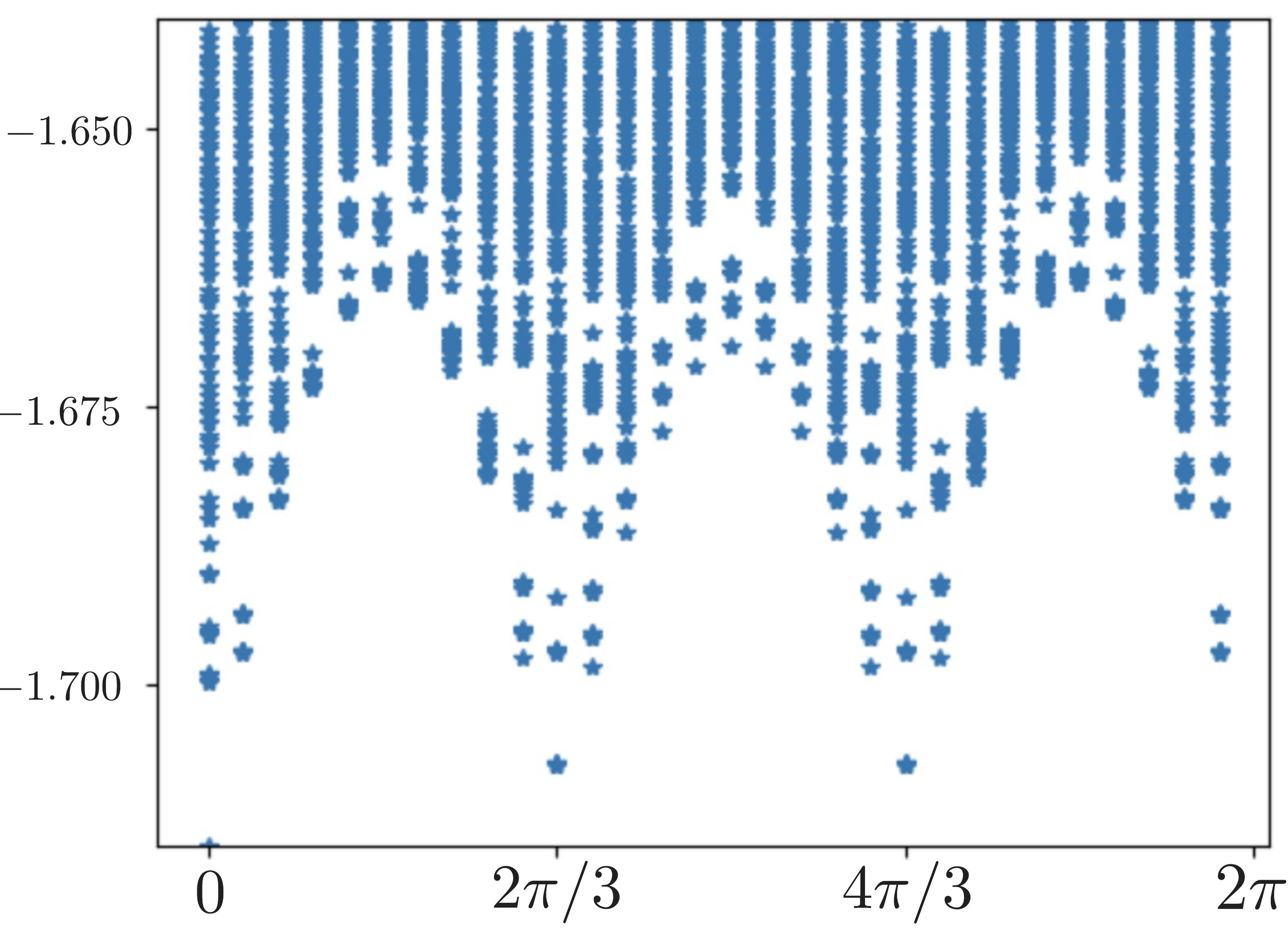} 
\caption{Spectrum (energy per-site) as a function of momentum at $J_x=J_z=1$ and $J_y=J_w=0$. Here we use periodic boundary conditions at $N=30$. Note that the gap closes around $k=0,\pm 2\pi/3$.}
\label{spec.pdf}
\end{figure}

We further numerically calculate the low energy spectrum using the periodic uniform Matrix Product States (puMPS)\cite{pumps1,pumps2}. Effectiveness of puMPS in extracting the spectrum of gapless spin chains was recently demonstrated in Refs.~\onlinecite{pumps3,pumps4}. We use the puMPS software package available in Ref.~\onlinecite{pumps5}. Sample results are shown in Fig.~\ref{spec.pdf}. In principle these results can be used to extract the entire conformal spectrum. However, the presence of logarithmic corrections\cite{ryu1,Affleck_1989} makes extracting the conformal data difficult. Nevertheless, we can verify numerically that the low-lying spectrum of scaling dimensions changes substantially as $\lambda$ is tuned.

Therefore, despite having the same central charge, the CFTs that emerge from different values of $\lambda$ have manifestly different spectra/scaling dimensions and are therefore not identical. They are adiabatically connected to $SU(3)_1$, and so they can be described by marginal deformations of $SU(3)_1$. Interestingly, all such CFTs have an emergent $U(1)^2$ symmetry, which implies that even though the microscopic symmetry is broken down to the discrete group $\mathbb{Z}_3\times \mathbb{Z}_3$, in the low energy limit the system hosts two \textit{emergent} conserved $U(1)$ charges (isospin and hyper-charge) throughout the critical phase.

From the puMPS results, we further observe that throughout the gapless phase of Eq.\eqref{h3}, there are three soft modes present at $k=0,\pm 2\pi/3$. This provides evidence that the microscopic translation symmetry acts as an internal $\mathbb{Z}_3$ in the low energy effective theory and is in agreement with the bosonization results derived below. 

\subsection{Two-component Luttinger liquid theory}
We now proceed to the explicit form of the low energy effective action for the gapless phase of Eq.\eqref{h3}. The most general form of the fixed point Lagrangian density of a
$c=2$ CFT that is smoothly connected to the $SU(3)_1$ WZW theory is given by
\begin{align}\label{string2}
  \mathcal{L} =&\frac{1}{2\pi}\Big[ G_{ij} \Big(\partial_t \varphi^i(x,t) \partial_t \varphi^j(x,t) - \partial_x \varphi^i(x,t) \partial_x \varphi^j(x,t) \Big)  \\ \newline \nonumber
  &+B_{ij}  \Big(\partial_t \varphi^i(x,t) \partial_x \varphi^j(x,t) - \partial_x \varphi^i(x,t) \partial_t \varphi^j(x,t) \Big) \Big] ,
 \end{align}
where $ \varphi^i \sim \varphi^i+2\pi$. We consider space to be a circle, so $x \sim x + 2\pi$. $G$ and $B$ are $2\times2$ symmetric and anti-symmetric matrices, respectively. We note that a symmetric tensor contribution to $B$ violates time-reversal and inversion symmetry and consequently is not considered; as we further discuss below, time-reversal and inversion each fix the anti-symmetric $B$ matrix to a fixed value, which can be non-zero due to $T$-duality. Furthermore, the on-site $\mathbb{Z}_3 \times \mathbb{Z}_3$ symmetry fixes $G$ up to an overall factor; any additional term $\propto G_{ij} \partial_x \varphi^i \partial_x \varphi^j$ can then be absorbed into a renormalization of $G_{ij}$ by rescaling space and time accordingly to change the overall velocity of excitations. Interestingly, the symmetries have enforced that the Lagrangian be conformally invariant; therefore the velocity of the two modes, for example, are equal.

This action gives the canonical momentum density,
 \begin{align}
 \pi^i(x,t) = \frac{\delta \mathcal{L}}{\delta \partial_t \varphi^i} = \frac{1}{2\pi}(2G_{ij} \partial_t \varphi^j(x,t) +  2B_{ij} \partial_x \varphi^j(x,t)).
 \end{align}
 To diagonalize the theory we mode expand the fields as
 \begin{align}
\varphi^i(x,t)=\varphi_0^i+ v_0^i t + m^i x + \sum_{l\neq 0} \varphi^i_l(t) e^{ilx},
 \end{align}
 with  $m^i,l \in \mathbb{Z}$. Here $\varphi_0^i$ and $v^i_0 $ are constants, independent of $x$ and $t$.

 The compactness of $\varphi$ implies that the momentum zero mode is quantized to integer values:
 \begin{align}
 p_0^i=\int_0^{2\pi} dx \pi^i=n^i \in \mathbb{Z}
   \end{align}
 Using this relation we find
 \begin{align}
v_0^i=\frac{1}{2}G^{-1}_{ij} n^j - G^{-1}_{ij} B_{jk} m^k.
 \end{align} 
 
 The spectrum of this model is given by a zero mode Hamiltonian, which gives the scaling dimensions of the conformal primaries, 
 combined with additional harmonic oscillator modes on top of them. Here we are only interested in the conformal primaries/scaling dimensions and
 ergo we safely set all harmonic modes $\varphi_l=0$. 
 
 The zero mode Hamiltonian can be written in terms of the integer vectors $n^T = (n^1, n^2)$ and $m^T = (m^1, m^2)$:
 \begin{align}\label{0m}
 H_0=& \frac{1}{2}\left(\begin{array}{cc} n^T & m^T\end{array}\right) \left(\begin{array}{cc} \frac{1}{2}G^{-1} & -G^{-1}B \\ BG^{-1} & 2(G-BG^{-1}B)\end{array}\right)  \left(\begin{array}{c} n \\ m\end{array}\right)  \\ \newline \nonumber
=& \frac{1}{2}\left(\begin{array}{cc} n^T & m^T\end{array}\right)  \mathcal{G}  \left(\begin{array}{c} n \\ m\end{array}\right),
 \end{align}
and
  \begin{align}
  \mathcal{G}^{-1}= \left(\begin{array}{cc}   2(G-BG^{-1}B) & BG^{-1} \\  -G^{-1}B &\frac{1}{2}G^{-1} \end{array}\right) .
  \end{align} 
  
  The Hamiltonian above is manifestly invariant under $(n,m)\rightarrow (m,n)$ and $\mathcal{G}\rightarrow\mathcal{G}^{-1}$. Moreover, discrete shifts $B\rightarrow B+\frac{1}{2}N$ and
  $(n,m) \rightarrow (n + Nm, m)$, where $N$ is an anti-symmetric integer matrix, also leave the spectrum invariant. These symmetries together generate the non-abelian T-duality group $O(3,3,\mathbb{Z})$ which generalizes the abelian $R\rightarrow \frac{1}{2R}$ duality of the $c=1$ Luttinger liquids\cite{giveon1,becker}. 

To see the decomposition of the Hamiltonian into left and right moving parts, it is useful to write the field $\varphi^i$ in terms of holomorphic andanti-holomorphic parts,
\begin{align}
\varphi^i(x,t)=\varphi_{L}^i(x+ t)+\varphi_{R}^i(x- t).
\end{align}
For the zero modes we have,
\begin{align}
&\varphi_{0,L}^i(x+ t)=\varphi_{0,L}^i+ p^i_{L} (x+ t) \\ \newline \nonumber
&\varphi_{0,R}^i(x- t)=\varphi_{0,R}^i+ p^i_{R} (x- t).
\end{align}
 Here we have defined $p^i_{L/R}$,
\begin{align}
p^i_L+p^i_R=m^i ~~\text{and}~~ p^i_L-p^i_R=v_0^i .
\end{align}
These variables give another form of the zero mode Hamiltonian,
\begin{align}\label{hlr}
 \frac{1}{2}H_0=p^i_L G_{ij} p^j_L +p^i_R G_{ij} p^j_R.
\end{align}
Eq. \ref{hlr} can be used to extract the left/right scaling dimension of primary operators.

The conventional ($2\pi$ periodic) Luttinger variables can be defined as
\begin{align}\label{lutt}
&\phi^i(x)=\varphi^i(x) = \varphi^i_L(x)+\varphi^i_R(x) \\ \newline \nonumber
&\theta^i(x)=2(G+B)_{ij}\varphi^j_L(x)-2(G-B)_{ij}\varphi^j_R(x).
\end{align}
Note that these operators are conjugate to each other, that is,
\begin{align}
\frac{1}{2\pi}\partial_x \theta^i = \pi^i.
\end{align}
  The Luttinger variables are defined such that,
\begin{align}\label{phit}
&\phi^i(x+2\pi)=\phi^i(x)+2\pi m^i \\ \newline \nonumber
&\theta^i(x+2\pi)=\theta^i(x)+2\pi n^i .
\end{align}

At the $SU(3)_1$ symmetric point we can choose\cite{giveon1},
\begin{align}
G^{SU(3)}=\frac{1}{4}\left(\begin{array}{cc} 2 & 1 \\ 1 & 2 \end{array}\right) ~ \text{and} ~ B^{SU(3)}=\frac{1}{4}\left(\begin{array}{cc} 0 & -1 \\ 1 & 0 \end{array}\right).
\end{align}

\subsection{Symmetry actions in Luttinger liquid theory}

To obtain the microscopic action of the symmetries we follow a method similar to Refs.~\onlinecite{affleck,affleck2,affleck3}. We assume that the $SU(3)$ invariant spin chain can be obtained from a large $U$ limit of a $1/3$ filled $SU(3)$ Hubbard model and that the symmetry actions can therefore be read off from (perturbative) Abelian bosonization of the fermions of the Hubbard model. We then use the same symmetry actions throughout the moduli space of the $c = 2$ Luttinger liquid. Details of this procedure are given in the Appendix~~\ref{spm}. Here we only quote the results. 

The $\mathbb{Z}_n$ generator $X$ acts as,
\begin{align}\label{eqx}
  X:~~&\phi^i \rightarrow M_{ij} \phi^j
\nonumber \\
      &\theta^i \rightarrow (M^T)^{-1}_{ij} \theta^j ,
\end{align}
for $M=\left(\begin{array}{cc} -1 & -1 \\ 1 & 0\end{array}\right)$.
Alternatively, we can define an auxiliary field $\varphi^3_{L/R}$ such that
\begin{align}
\varphi^1_L+\varphi^2_L+\varphi^3_L=0,
\end{align}
and a similar relation for $\varphi_R$. Then $X$ acts as a cyclic permutation $\varphi^3\rightarrow\varphi^2\rightarrow\varphi^1$.

The action of the $\mathbb{Z}_3$ symmetry generator $Z$ is
\begin{align}\label{eqz}
  Z:~~&\theta^\alpha \rightarrow \theta^\alpha +  2\pi \alpha/3
\nonumber \\
 & \phi^i \rightarrow  \phi^i .
\end{align}
As discussed earlier, at special points the $\mathbb{Z}_3 \times \mathbb{Z}_3$ is enhanced to  $\mathbb{Z}_3 \ltimes U(1)^2$. This microscopic $U(1)^2$ symmetry acts as,
\begin{align}
 U(1)^2:~~&\theta^\alpha \rightarrow \theta^\alpha +  c^\alpha\nonumber \\
 & \phi^i \rightarrow  \phi^i,
\end{align}
where $c^\alpha$ is an arbitrary real number. 

To be consistent with the $\mathbb{Z}_3 \times \mathbb{Z}_3$ symmetry, the $G$ and $B$ matrices must satisfy
\begin{align}
M^T G M=G, \;\; M^T B M=B .
\end{align}
It is straightforward to check that in fact the most general $G$ and $B$ matrices consistent with the $\mathbb{Z}_3 \times \mathbb{Z}_3$ symmetry
satisfy $G\propto G^{SU(3)}$ and $B \propto B^{SU(3)}$. This means that by requiring  $\mathbb{Z}_3\times \mathbb{Z}_3$ symmetry the $G,B$ matrices can be written as
\begin{align}
  G &=\frac{g}{4}\left(\begin{array}{cc} 2 & 1 \\ 1 & 2 \end{array}\right)
                                                    \nonumber \\
  B &=\frac{b}{4}\left(\begin{array}{cc} 0 & -1 \\ 1 & 0 \end{array}\right).
\end{align}

Translation by one site $T_x$ acts as,
\begin{align}\label{eqtx}
  T_x: ~~ & \theta^\alpha\rightarrow \theta^\alpha+ 2\pi \alpha/3
\nonumber \\
 & \phi^\alpha\rightarrow \phi^\alpha+2\pi/3.
\end{align} 
It is now manifest that translation by a single site acts as an emergent $\mathbb{Z}_3$ internal symmetry in the low energy theory.
This $\mathbb{Z}_3$ symmetry associated with translations may itself be anomalous\cite{Kikuchi,GEPNER1,oshikawa1,oshikawa2},
although we do not investigate this here. 

We now discuss the discrete $\mathcal{C},P,\Theta$ symmetries. Inversion $P$ acts like ($\varphi_L\rightarrow -\varphi_R$),
 \begin{align}\label{eqp}
   P: ~~ &\theta^i\rightarrow \theta^i - 4B^{SU(3)}_{ij} \phi^j
     \nonumber \\
         &\phi^i\rightarrow -\phi^i .
 \end{align}
 Time-reversal (complex conjugation in $Z$ basis) acts as ($\varphi_L\rightarrow \varphi_R$),
  \begin{align}\label{eqp2}
    \Theta: ~~ &\theta^i\rightarrow -\theta^i + 4B^{SU(3)}_{ij} \phi^j
    \nonumber \\
  &  \phi^i\rightarrow \phi^i .
  \end{align}
  Finally, charge-conjugation acts as ($\varphi^1\rightarrow \varphi^2$),
 \begin{align}\label{eqp3}
\mathcal{C}: ~~ &(\theta^1, \theta^2) \rightarrow (\theta^2 - \phi^1, \theta^1 + \phi^2) \nonumber \\
& (\phi^1, \phi^2) \rightarrow (\phi^2,\phi^1) .
 \end{align}

The charge-conjugation, inversion and time-reversal symmetries flip the sign of the second term in Eq.\eqref{string2}, effectively taking $B\rightarrow -B$. At the $SU(3)$ point $\mathcal{C},P,\Theta$ are good symmetries, since under $B\rightarrow -B$, the spectrum is invariant ($4B^{SU(3)}$ is an integer valued matrix). As we deviate from the $SU(3)$ point, these symmetries force the $B$ matrix to be fixed $B=B^{SU(3)}$  (i.e. $b=1$). Therefore, the most general $G$ and $B$ matrices consistent with $\mathbb{Z}_3\times \mathbb{Z}_3$ \textit{and} $\mathcal{C},P,\Theta$ that are adiabatically connected to the $SU(3)$ point are,
\begin{align}
G=\frac{g}{4}\left(\begin{array}{cc} 2 & 1 \\ 1 & 2 \end{array}\right) ~ \text{and} ~ B=\frac{1}{4}\left(\begin{array}{cc} 0 & -1 \\ 1 & 0 \end{array}\right).
\end{align}
The moduli space of $\mathbb{Z}_3\times \mathbb{Z}_3$ \textit{and} $\mathcal{C},P,\Theta$ symmetric $c=2$ Luttinger liquids is therefore \textit{one} dimensional and characterized by a single real parameter $g$. We remark that in principle it is possible to also have a symmetric theory with a vanishing $B$ matrix. However, this theory is not adiabatically connected to the $SU(3)$ point and we do not consider it here. It is remarkable that the  $\mathbb{Z}_3\times \mathbb{Z}_3$ and $\mathcal{C},P,\Theta$ imply that the Luttinger liquid is \it conformally invariant \rm, since multi-component Luttinger liquids generically need not be conformally invariant. 

Finally, we note that since the charge conjugation symmetry interchanges $\phi^1$ and $\phi^2$,  it alone is sufficient to force the velocity of the two modes to be the same.

\subsection{LSM anomaly of two-component Luttinger liquid}

Similar to the $\mathbb{Z}_2 \times \mathbb{Z}_2$ case, to directly see the mixed LSM anomaly, we need to insert a unit of translational symmetry flux through the system. Inserting a unit of $T_x$ flux changes Eq.\eqref{phit} according to,
 \begin{align}\label{phit1}
&\phi^\alpha(x+2\pi)=\phi^\alpha(x)+2\pi m^\alpha +  2\pi/3\\ \newline \nonumber
&\theta^\alpha(x+2\pi)=\theta^\alpha(x)+2\pi n^\alpha + 2\pi\alpha/3.
\end{align}
This can be absorbed into a shift of $n^i$ and $m^i$,
\begin{align}
m^\alpha\rightarrow m^\alpha + \frac{1}{3} ~~ \text{and} ~~  n^\alpha\rightarrow n^\alpha +\frac{\alpha}{3}.
\end{align}

Plugging $n,m$ into the zero mode Hamiltonian Eq.\eqref{0m} gives the zero mode energies of the spin chain with $3n+1$ sites.  It is easy to verify that the six states 
\begin{align}
&\{n_1=+1/3,n_2=-1/3,m_1=-2/3,m_2=+1/3\},\\ \newline \nonumber
&\{n_1=+1/3,n_2=-1/3,m_1=+1/3,m_2=-2/3\},\\ \newline \nonumber
&\{n_1=-2/3,n_2=-1/3,m_1=+1/3,m_2=+1/3\},\\ \newline \nonumber
&\{n_1=+1/3,n_2=+2/3,m_1=+1/3,m_2=+1/3\},\\ \newline \nonumber
&\{n_1=-2/3,n_2=-1/3,m_1=-2/3,m_2=+1/3\},\\ \newline \nonumber
&\{n_1=+1/3,n_2=+2/3,m_1=+1/3,m_2=-2/3\},
\end{align}
are degenerate ground states at energy $E=1/3$, which matches the known results for $SU(3)_1$ WZW\cite{su3dmrg}. This degeneracy is expected to break to two sets of three-fold degenerate states in finite-size systems by irrelevant or marginally irrelevant operators that break $\frac{SU(3)_L\times SU(3)_R}{\mathbb{Z}_3} \rightarrow PSU(3)$. This is a direct manifestation of the mixed LSM anomaly. A similar relation also holds for $3n+2$ sites.

Moreover, it should be emphasized that the three-fold degeneracy is guaranteed as long as $G\propto G_{SU(3)}$. That is, all $\mathbb{Z}_3\times \mathbb{Z}_3$ symmetric points have the same mixed LSM anomaly. To see this note that for $G\propto G_{SU(3)}$, the zero mode Hamiltonian Eq.\eqref{0m} is invariant under the $\mathbb{Z}_3$ action,
\begin{align}\label{xz3}
 \left(\begin{array}{c} n \\ m\end{array}\right) \rightarrow \left(\begin{array}{cc} (M^T)^{-1} & 0 \\ 0 & M\end{array}\right)  \left(\begin{array}{c} n \\ m\end{array}\right).
\end{align}
Unless the total charge $(n ,m)=0$, the action above generates three different degenerate ground states. Therefore, when the zero charge state is forbidden by having $3n\pm1$ sites on the lattice or equivalently having non-trivial $T_x$ flux in the field theory, the ground state is at least three fold degenerate. We note that this calculation is independent of the value of $B$, as long as the $\mathbb{Z}_3 \times \mathbb{Z}_3 \times \mathbb{Z}_{\text{trans}}$ symmetry is preserved.

\section{Stability of two-component Luttinger liquid}
\label{stabilitySec}  

The most generic perturbations that can be added to the fixed point action and that can potentially destabilize the critical phase consist of superpositions of vertex operators
\begin{align}
\mathcal{O}_{mn} = \exp(i (m^i\theta^i+n^i\phi^i)) .
\end{align}
Note that $P\Theta$ is an anti-unitary operator that takes $\mathcal{O}_{mn} \rightarrow \mathcal{O}_{mn}$. Therefore
$P\Theta$ allows only the cosine combinations to appear
\begin{align}
\mathcal{O}_{mn} + \mathcal{O}_{mn}^\dagger \propto \cos( m^i\theta^i+n^i\phi^i ).
  \end{align}
The scaling dimension of the operator above is given by Eq.\eqref{0m} (right/left scaling dimension are also given in Eq.\eqref{hlr}).

The most relevant operators that are consistent with all of the symmetries are
 \begin{align}\label{pert}
   V_1 = &\cos (\phi^1-\phi^2)+\cos (2\phi^2+\phi^1)+\cos (2\phi^1+\phi^2),\\ \newline \nonumber
   V_2 = &\cos (\theta^1+\theta^2)+\cos (2\theta^2-\theta^1)+\cos (2\theta^1-\theta^2)+ \\ \newline \nonumber
 &\cos (\phi^1-\phi^2-\theta^1-\theta^2)+\cos (\phi^1+2\phi^2+2\theta^1-\theta^2) \\ \newline \nonumber
         &+\cos (2\phi^1+\phi^2+\theta^1-2\theta^2) ,
   \nonumber \\
    V_3 = &\cos [2(\theta^1+\theta^2)+(\phi^2-\phi^1))]\\ \newline \nonumber
&+\text{terms related by the action of $X$ symmetry}.
\end{align}
$V_1$ has scaling dimension $2/g$ and is relevant for $g>1$. $V_2$ has scaling dimension $\frac{1+3g^2}{2g}$ and is relevant for $\frac{1}{3}<g<1$. At the $PSU(3)$ symmetric point $g=1$, $V_1$ and $V_2$ are marginal. $V_3$ has scaling dimension $6g$ and is therefore relevant for $0<g<1/3$, where $V_1$ and $V_2$ are irrelevant. At the $g=1$ point, $V_3$ has scaling dimension $6$ and is strongly irrelevant. 
 
Therefore, away from the point $g=1$, a relevant symmetry-allowed operator is always present. This naively suggests that unless additional symmetries can be enforced that could rule out these operators, the critical phase of these systems should be unstable. If the RG flow is to a gapped state, on general grounds one also expects an energy gap of order 1 in units set by the microscopic energy scales, unless the system is fine-tuned to be close to the gapless $SU(3)$ point. However this deduction appears to be in direct contradiction with the numerical results presented here and obtained by other groups\cite{ryu1,ryu2}.

The numerical results strongly suggest that these critical phases are in fact stable -- at least to an excellent approximation -- for a wide range of microscopic parameters of the nearest neighbor Hamiltonian. For example, for the quantum torus chain Hamiltonian, Eq. \ref{qtc}, for a large region of the phase diagram we find a gapless phase with $c = 2$ (see Fig. \ref{fig1prime}). Alternatively, starting at the $PSU(3)$ invariant point, we can increase $J_z$ to be arbitrarily large, with the gapless nature of the system appearing to persist throughout. These are parameter ranges of order infinity in dimensionless units, and therefore this behavior cannot be understood by assuming that the perturbations are only weakly relevant so that the numerically accessible system sizes are smaller than the correlation length, unless one also has the unnatural scenario where the bare coupling for the relevant operators always remains small throughout these large changes of the microscopic parameters. 

In the next subsections, we provide evidence that the $g > 1$ regime of the Luttinger liquid can be accessed when $J_w=J_x=J_y=J$ and $J_z>|J|$, and that the $g < 1$ regime can be accessed when $J_w=J_x=J_y=J$ and $-|J| < J_z < |J|$. Note that in these parameter regimes, the $PSU(3)$ symmetry is broken to $ \mathbb{Z}_3\ltimes U(1)^2$. We thus provide numerical and analytical evidence that despite the apparent existence of symmetry-allowed relevant operators in the field theory, the critical phase is still stable, at least to an excellent approximation, for a wide range of parameters of the microscopic Hamiltonian. At its core, this surprising result can be related to the fact that the parameters of the spin chain Hamiltonian are associated with microscopic terms that are frustrated, in the sense that the ground state subspace associated with them is exponentially large.

\subsection{Stability of the critical phase for $g>1$}

In the regime $g>1$ the most relevant operator is given by,
 \begin{align}\label{oppert}
V_1 = \cos (\phi^1-\phi^2)+\cos (2\phi^2+\phi^1)+\cos (2\phi^1+\phi^2).
\end{align}
The other operators introduced in Eq.\eqref{pert} are irrelevant in this regime.

If $V_1$ enters the Hamiltonian with a negative sign, at strong coupling it can pin both its arguments $\phi_1=\phi_2=0,\pm \frac{2\pi}{3}$, giving rise to gapped phase with threefold ground state degeneracy. However, if $V_1$ comes with a positive sign, all three of the cosine operators cannot be simultaneously minimized. Yet, there are $6$ minima $\phi_1=0,\pm \frac{2\pi}{3}, \phi_2=\phi_1\pm \frac{2\pi}{3}$, favoring a period $3$ anti-ferromagnetic pattern $\cdots g_1g_2g_3g_1g_2g_3\cdots$ or $\cdots g_3g_2g_1g_3g_2g_1\cdots$. Therefore it is natural to expect that if the CFT is perturbed by this relevant operator, the RG flow will be towards a gapped symmetry-breaking phase. 

Below we provide a series of arguments to support the claim that the regime $g > 1$ can be accessed in the spin chain Hamiltonian by setting $J_w=J_x=J_y=J$ and $J_z>|J|$. This is surprising since the numerics indicate that this region of the phase diagram is gapless, despite the fact that $V_1$ is relevant and symmetry-allowed in this regime. 
\begin{figure*}[t]
    \centering
    \begin{subfigure}[t]{0.5\textwidth}
        \centering
\includegraphics[width=\columnwidth,keepaspectratio]{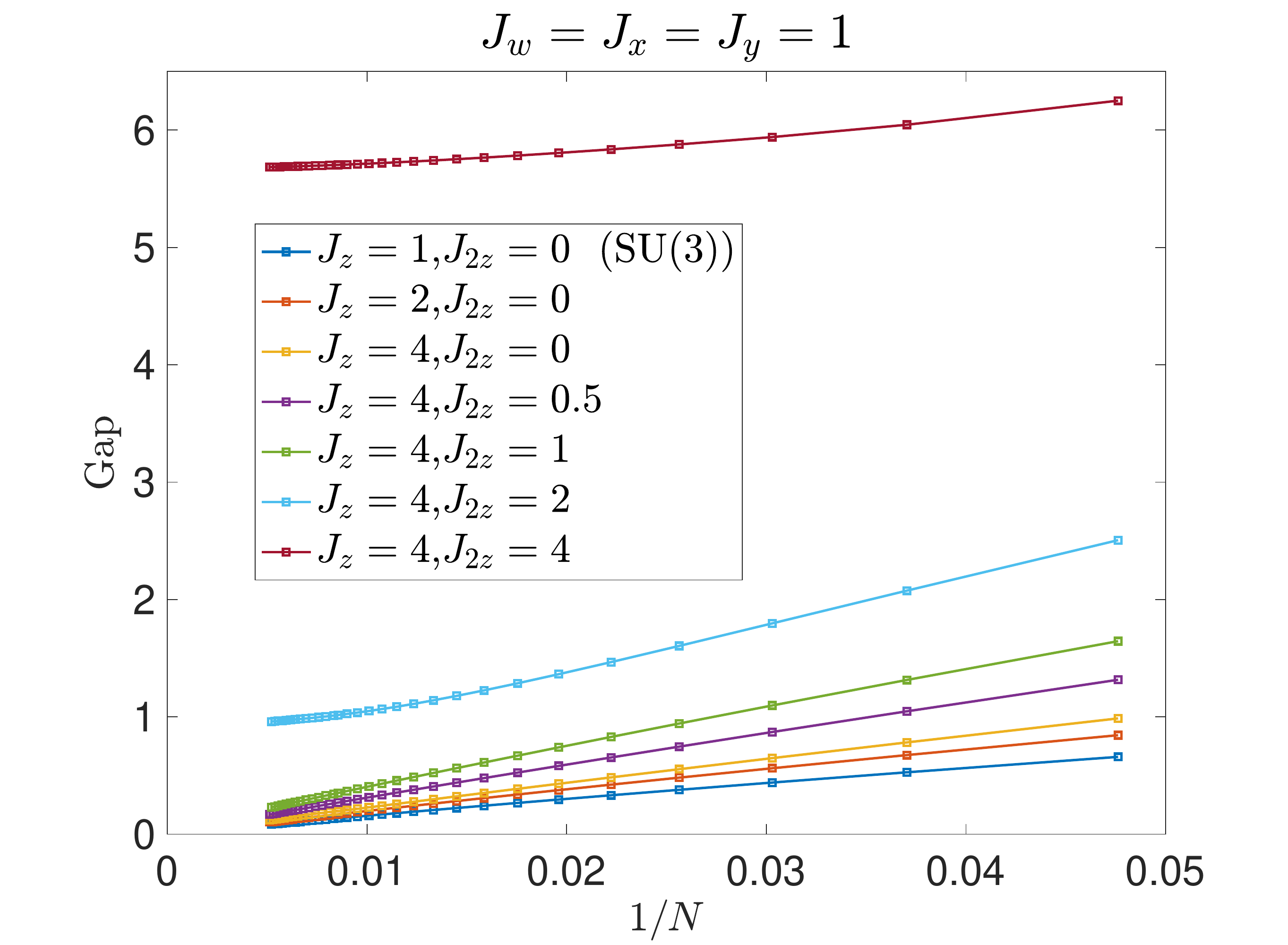} 
\caption{}
\label{g1}
    \end{subfigure}%
    ~ 
    \begin{subfigure}[t]{0.5\textwidth}
        \centering
        \includegraphics[width=1.02\columnwidth,keepaspectratio]{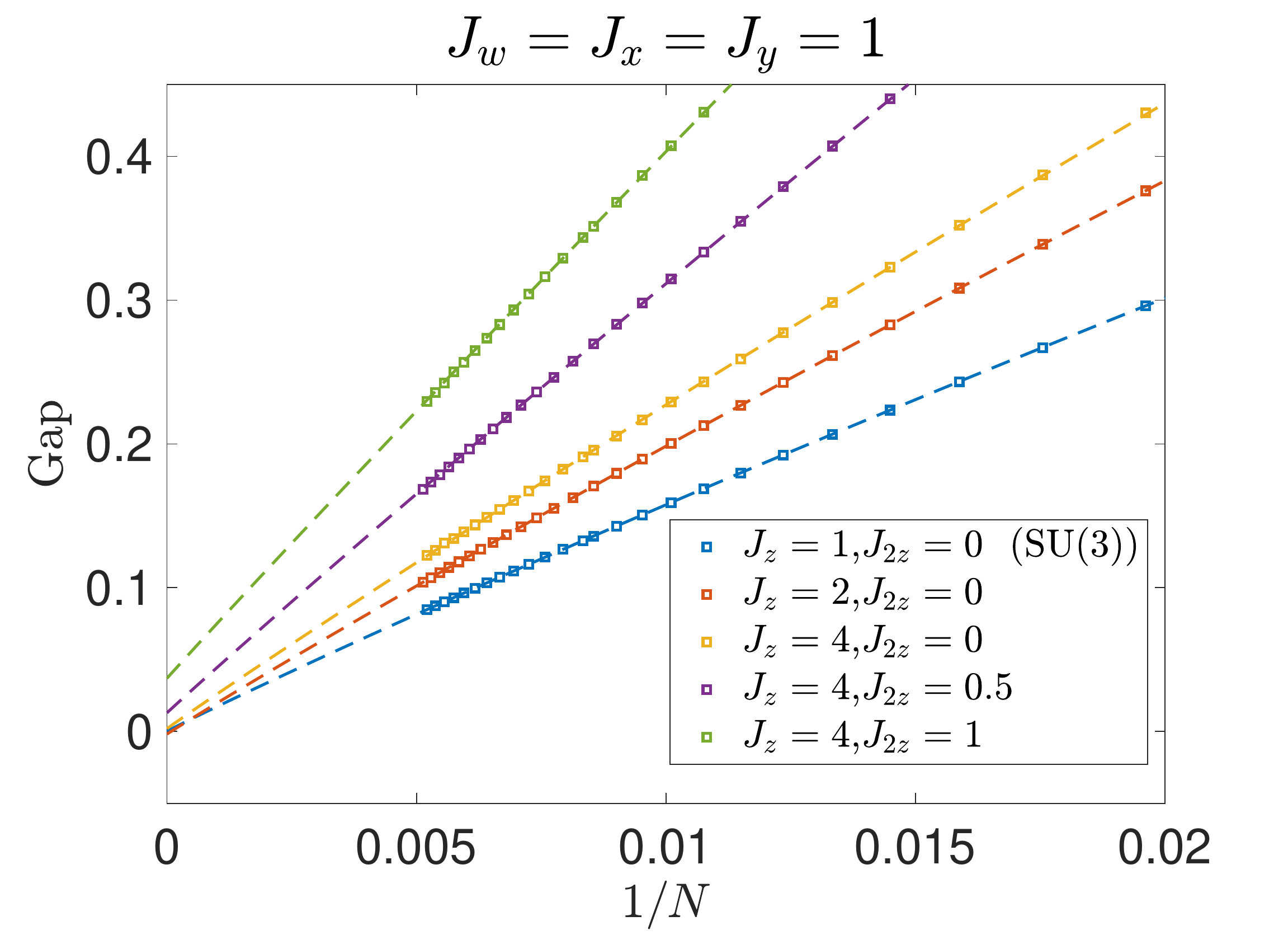} 
\caption{}
\label{g2}
    \end{subfigure}
    \caption{Finite size scaling of the energy difference between $S_z=0$ and $S_z =1$ sectors in the $J_z>|J|$ gapless regime at $J=J_w=J_x=J_y=1$. We use open boundary condition and consider systems sizes up to $N=195$. We find that a nonzero second nearest neighbor term induces a finite gap in the thermodynamic limit. Subfigure (b) is the zoomed-in version of subfigure (a).}
\end{figure*}

First, that $g > 1$ can be accessed in the above stated regime can be seen by considering the $\sum_i Z_i Z_{i+1}^\dagger+Z_i^\dagger Z_{i+1}$ term in the spin chain Hamiltonian Eq.\eqref{h3}, which can be expanded in terms of the fields at long wavelengths as
\begin{align}\label{fttt}
\sum_i Z_i Z_{i+1}^\dagger+Z_i^\dagger Z_{i+1} = \int dx [a G_{ij} \partial_x \phi^i \partial_x \phi^j  + b V_1 + \cdots],
\end{align}
where $a$ and $b$ are in general functions of the couplings $J_i$ and the $\cdots$ indicate less relevant operators. Note that the above expansion can be deduced purely on the basis of symmetry principles; $V_2$ and $V_3$ do not appear in the expansion above as they break the $U(1)^2$ symmetry. $\partial_x \theta$ dependent terms also do not appear as they violate the local charge conservation symmetry (onsite charge is conserved not just global charge) $\theta^\alpha(x)\rightarrow \theta^\alpha(x)+f^\alpha(x)$ respected by $\sum_i Z_i Z_{i+1}^\dagger+Z_i^\dagger Z_{i+1}$. One can check that a positive $a$ effectively increases $g$ while also renormalizing the velocity of the excitations, while negative $a$ decreases $g$. 

The bosonization results (see Appendix) imply that $a$ should be positive near the $SU(3)$ point, which therefore suggests that increasing $J_z$ should take the field theory to the $g > 1$ parameter regime. We note that the fact that $g > 1$ in the regime $J_w=J_x=J_y=J$ and $J_z>|J|$ is also in agreement with results of Ref.~\onlinecite{ryu2}, where the scaling dimensions were numerically calculated and it was argued that the system is described by the two-component Luttinger liquid with $g > 1$. However this is a highly non-trivial result, because Eq. \ref{fttt} would naively suggest that perturbing by a small positive $\delta H = \alpha \sum_i (Z_i Z_{i+1}^\dagger+Z_i^\dagger Z_{i+1})$ should trigger an RG flow controlled by the relevant operator $V_1$, and thus would gap the system. Ref.~\onlinecite{ryu2} did not consider the presence of this relevant operator. It is thus worth further confirming that in this regime, the spin chain is indeed described by $g > 1$, which we do in the subsequent subsections. 

If the spin chain is indeed described by $g > 1$ in the regime $J_w=J_x=J_y=J$ and $J_z>|J|$, the gaplessness (or near gaplessness) of the system for an essentially infinite range of $J_z/|J|$ is extremely surprising. One possibility is that $b \approx 0$ for a wide range of $J_i$, in Eq. \ref{fttt}. As discussed in the Appendix, the bosonization at the $SU(3)$ point is actually ambiguous, and there is a way of carrying out the bosonization in which $b = 0$ at the $SU(3)$ point. In principle there could be additional conserved quantities, or the model could even be integrable, away from the $SU(3)$ point which could potentially explain why $b = 0$ throughout a large region of the phase diagram. On the other hand, it could be the case that accidentally $b \approx 0$ for a large region of the phase diagram, although this is unnatural from the perspective of the field theory. 

Eq. \ref{fttt} also contains a tower of higher order irrelevant operators in the expansion and also breaks Lorentz symmetry. It is therefore not \it a priori \rm clear what the result of the RG flow would be when the system is perturbed in this direction, even for non-zero $b$. Additionally, because the perturbation breaks Lorentz invariance, the $c$-theorem cannot be used to anticipate that the central charge must necessarily decrease under the RG flow. Therefore, in principle the system can remain gapless and stay at the same central charge for non-zero $b$, although this would be surprising. 

We note that the microscopic operator $\sum_i (Z_i Z_{i+1}^\dagger+Z_i^\dagger Z_{i+1})$ is frustrated. The ground state of the $\sum_i (Z_i Z_{i+1}^\dagger+Z_i^\dagger Z_{i+1})$ term is given by all spin configurations $|\{g_i\}\rangle$ where no nearest neighbor spins are the same $g_i\neq g_{i+1}$. In the $n=2$ case this ground state subspace would have been two-fold degenerate. However, for $n=3$ this ground state subspace is exponentially large, as there are $d_g=3\times 2^{N-1}$ such states. The factor of $2^{N-1}$ in the degeneracy can be seen to arise from the existence of a $U(2)^{N-1}$ symmetry of the term $\sum_i (Z_i Z_{i+1}^\dagger+Z_i^\dagger Z_{i+1})$. One can verify that associated with each bond $(i,i+1)$, there is a $U(2)$ group of unitary transformations that commute with $\sum_i (Z_i Z_{i+1}^\dagger+Z_i^\dagger Z_{i+1})$. The unitaries associated with each bond are highly non-local in terms of the original spin variables. 

The analog of this lattice frustration is missing from the leading terms in the field theory expansion (Eq.\eqref{fttt}). It is clear then that the higher order terms in Eq. \eqref{fttt} are necessarily important in determining the effect of a perturbation by $(Z_i Z_{i+1}^\dagger+Z_i^\dagger Z_{i+1})$. Moreover, it is not clear what the consequences are of the $U(2)^{N-1}$ symmetry for the long wavelength expansion of $\sum_i (Z_i Z_{i+1}^\dagger+Z_i^\dagger Z_{i+1})$, and in particular whether it would force $b= 0$ in Eq. \ref{fttt}. 

To help establish that we are in the $g > 1$ regime, below we provide numerical evidence that second neighbor perturbations and also inversion symmetry breaking perturbations induce perturbations to the field theory that are relevant for $g > 1$.

\subsubsection{Evidence for relevance of second neighbor perturbations}

Let us consider adding second nearest neighbor terms of the the form $\sum_iZ_i Z_{i+2}^\dagger+Z_i^\dagger Z_{i+2}$ to the Hamiltonian. In the continuum description we have,
 \begin{align}
  \sum_i[J_z(Z_i Z_{i+1}^\dagger+&Z_i^\dagger Z_{i+1}) +  J_{2z} (Z_i Z_{i+2}^\dagger+Z_i^\dagger Z_{i+2})]   \\ \newline \nonumber
& = \int dx [\tilde{a} G_{ij} \partial_x \phi^i \partial_x \phi^j + \tilde{b} V_1 + \cdots], 
\end{align}
where $J_z$, $J_{2z}$ are positive constants, and again $\tilde{a}$ and $\tilde{b}$ are in principle functions of the microscopic parameters $J_i$. We argue that $\tilde{a}$ is positive, which keeps us in the $g > 1$ regime, which is reproduced from the bosonization results of the appendix. The relevant operator $V_1$ (with positive $\tilde{b}$ as expected from the bosonization) again favors a period $3$ anti-ferromagnetic pattern $\cdots g_1g_2g_3g_1g_2g_3\cdots$ or $\cdots g_3g_2g_1g_3g_2g_1\cdots$. However, in this case, the microscopic term also favors the the same state. That is, the presence of the second nearest neighbor term removes the frustration and leads to an agreement between the microscopic and the continuum descriptions. In this case, we expect that a small perturbation involving $J_z$ and $J_{2z}$ together is expected to gap the system. 
\begin{figure*}[t]
    \centering
    \begin{subfigure}[t]{0.5\textwidth}
        \centering
        \vspace{0mm}
\includegraphics[width=\columnwidth,keepaspectratio]{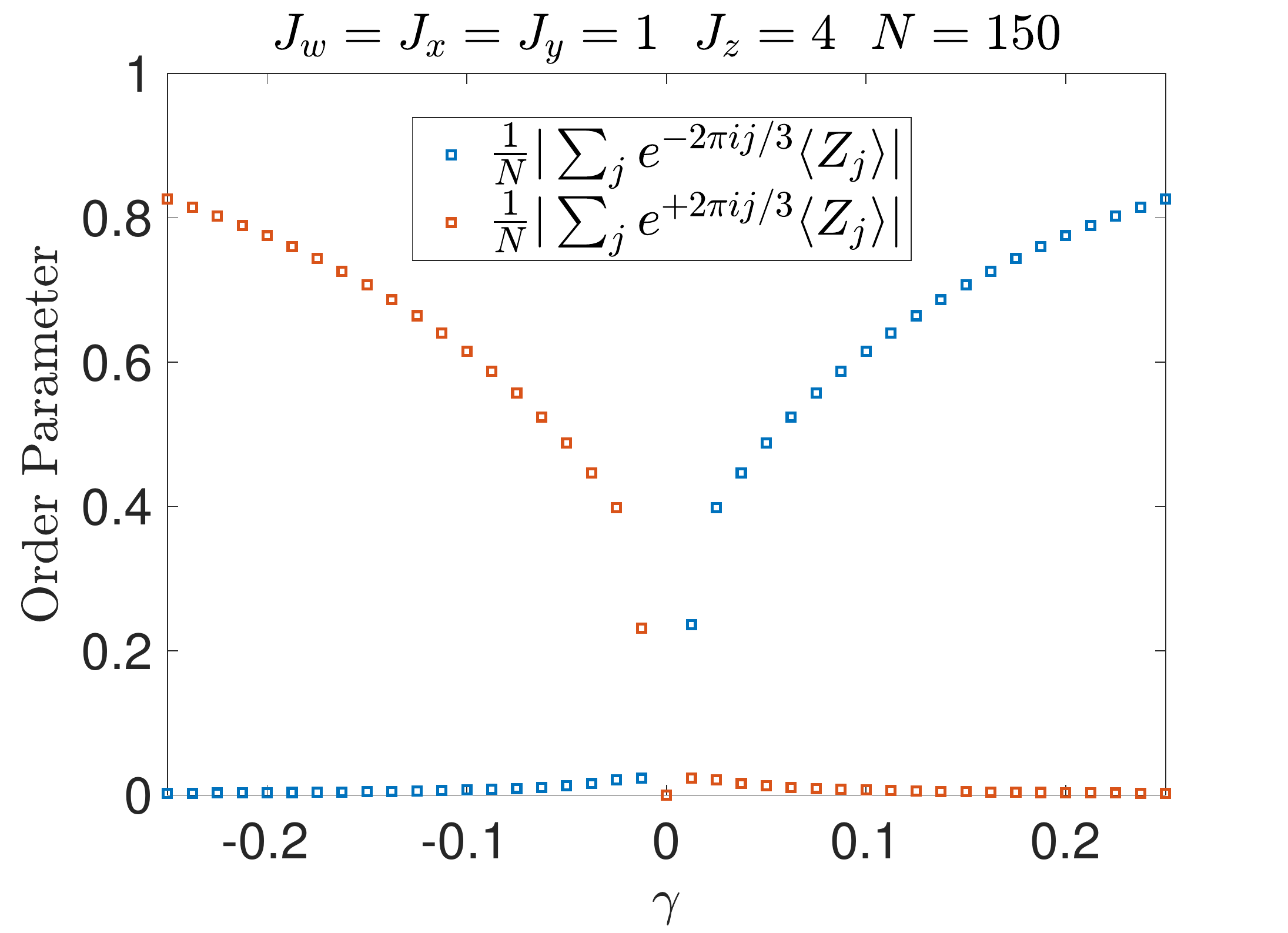} 
\caption{}
\label{fig4}
    \end{subfigure}%
    ~ 
    \begin{subfigure}[t]{0.5\textwidth}
        \centering
       \vspace{0mm}
\includegraphics[width=\columnwidth,keepaspectratio]{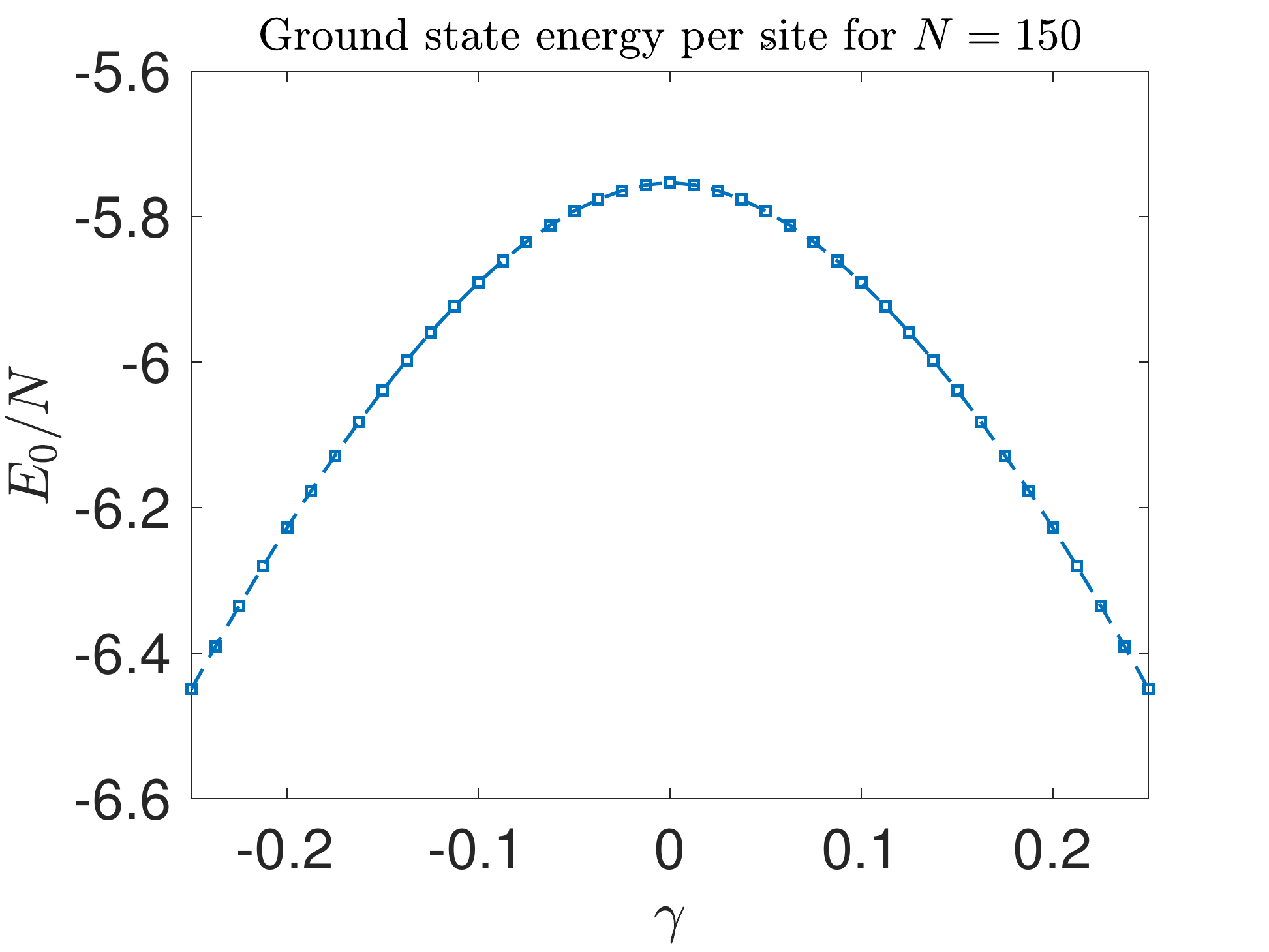} 
\caption{}
\label{fig5}
    \end{subfigure}
    \caption{Order parameter and ground state energy as function of $\gamma$ (defined in Eq.\eqref{numh}) at $J=J_w=J_x=J_y=1$ and $J_z=4$. Here we have used open boundary conditions and set the system size $N=150$. The results are consistent with a continuous phase transition.}
\end{figure*}
To check the above hypothesis numerically, we have simulated the Hamiltonian in presence of the second nearest neighbor term,
\begin{align}\label{hn1}
H=\sum_i \Big( &W_i W^\dagger_{i+1}+ X_i X^\dagger_{i+1} +  Y_i Y^\dagger_{i+1} \\ \newline \nonumber
&+J_z Z_i Z^\dagger_{i+1} +J_{2z} Z_i Z^\dagger_{i+2}  \Big)+h.c~.
\end{align}
Figs.~\ref{g1} and  ~\ref{g2} show the energy difference between $S_z=0$ and $S_z =1$ sectors in this regime. Consistent with our expectation, we find that (at least a strong enough) second nearest neighbor term induces a finite gap in the thermodynamic limit. By choosing a large enough $J_{2z}$, we can make this gap arbitrarily large.

We note that the gap of Eq. \ref{hn1} provides further evidence that the spin chain is in the regime $g > 1$. For $g < 1$, $V_1$ is irrelevant, and therefore one would not expect that adding a small second neighbor $J_{2z}$ term would gap the system.
\begin{figure*}[t]
    \centering
    \begin{subfigure}[t]{0.5\textwidth}
        \centering
        \vspace{0mm}
\includegraphics[width=0.93\columnwidth,keepaspectratio]{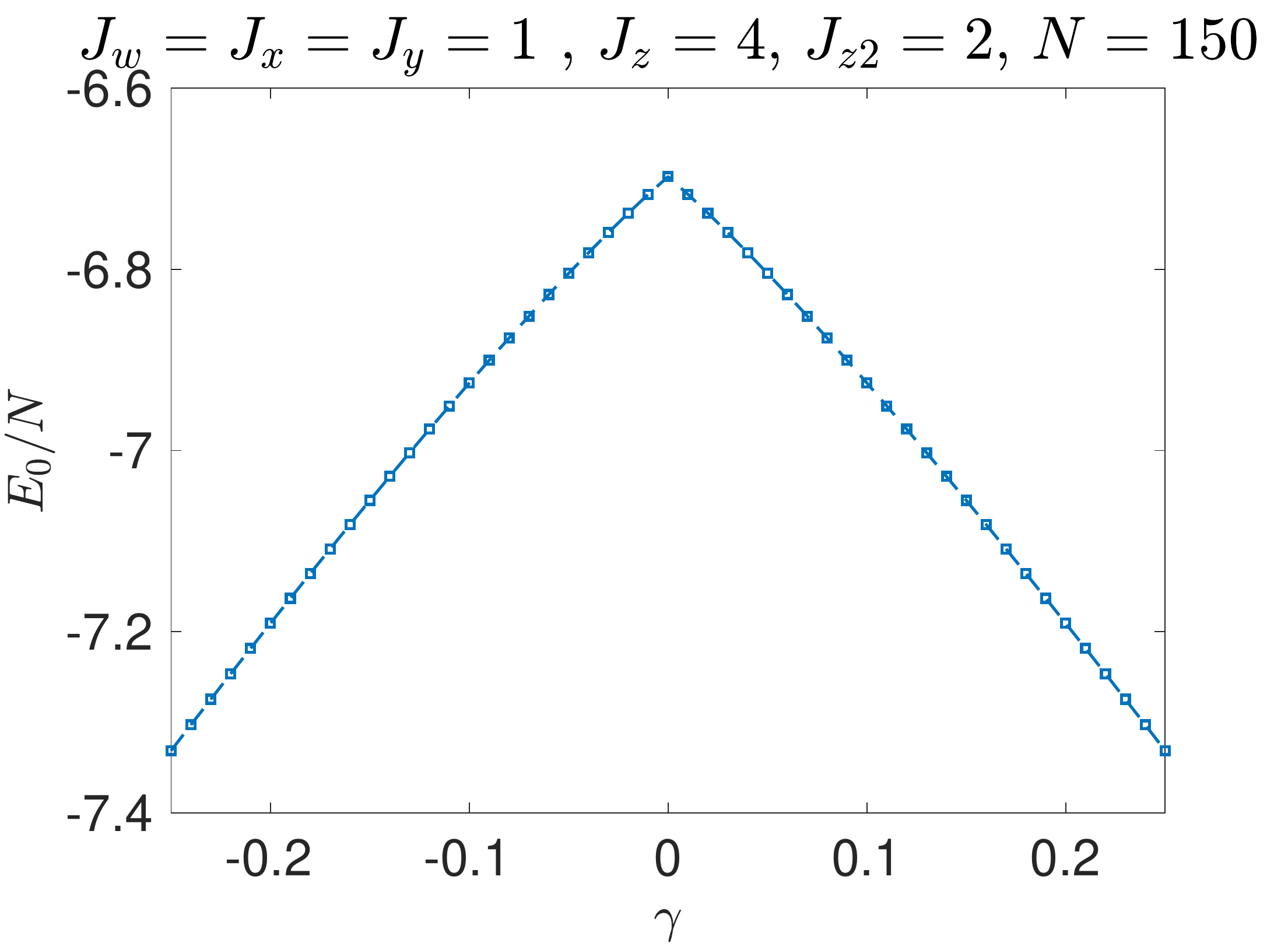} 
\caption{}
\label{g3}
    \end{subfigure}%
    ~ 
    \begin{subfigure}[t]{0.5\textwidth}
        \centering
       \vspace{0mm}
\includegraphics[width=\columnwidth,keepaspectratio]{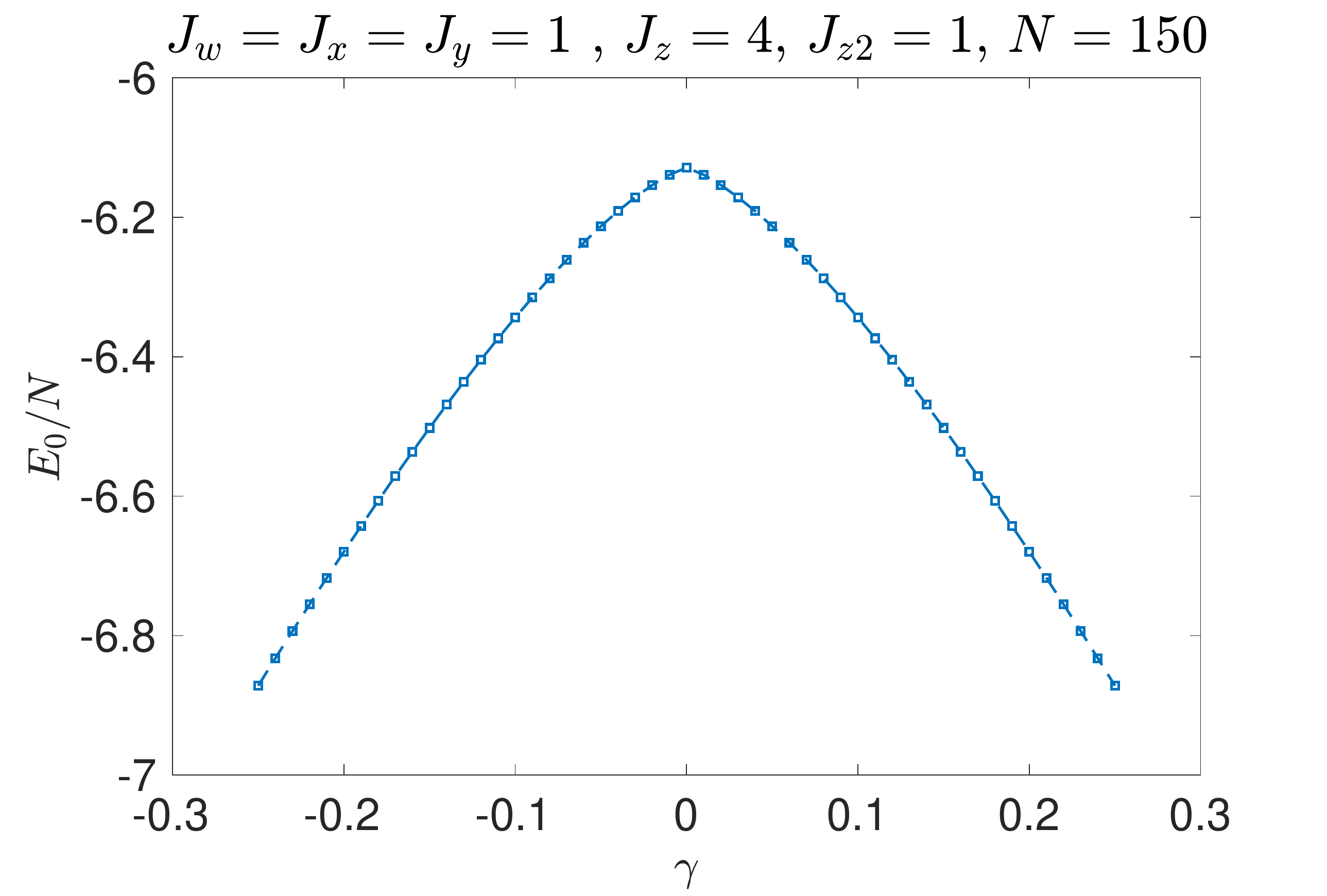} 
\caption{}
\label{g4}
    \end{subfigure}
    \caption{Ground state energy as function of $\gamma$ with nonzero second nearest neighbor terms $J_{2z}=2$ and $J_{2z}=1$ at $J=J_w=J_x=J_y=1$ and $J_z=4$. Here we have used open boundary conditions and set the system size $N=150$. The results are consistent with a first order phase transition}
\end{figure*}
\subsubsection{Relevant inversion and charge-conjugation symmetry-breaking perturbations from imaginary $J_z$}

Let us further consider adding a symmetry-breaking imaginary part to the $J_z$ term,
\begin{align}
 i \sum_i (Z_i Z_{i+1}^\dagger-Z_i^\dagger Z_{i+1}).
\end{align}
This term breaks both inversion and charge conjugation symmetries. As opposed to the term with real $J_z$, the ground state of this term favors a period $3$ anti-ferromagnetic pattern ($\cdots g_1g_2g_3g_1g_2g_3\cdots$ or $\cdots g_3g_2g_1g_3g_2g_1\cdots$ depending on the sign of the term above) and has no exponential ground state degeneracy. In the effective field theory, this term generates the operator
\begin{align}
  \label{V1tilde}
\tilde{V}_1 = \sin (\phi^1-\phi^2)+\sin (2\phi^2+\phi^1)+\sin (-2\phi^1-\phi^2).
\end{align}
Similar to $V_1$, $\tilde{V}_1$ also has scaling dimension $2/g$ and is therefore relevant in the $g>1$ regime. Since the corresponding microscopic term has no exponential degeneracy, we expect this term to induce a finite gap. To test this numerically, we have simulated the spin chain with Hamiltonian ($J_z>1$),
\begin{align}\label{numh}
H=\sum_i \Big( &W_i W^\dagger_{i+1}+ X_i X^\dagger_{i+1} +  Y_i Y^\dagger_{i+1} \\ \newline \nonumber
&+J_z(1+i \gamma ) Z_i Z^\dagger_{i+1}  \Big)+h.c~.
\end{align}

\begin{figure}[t]
\centering
	\vspace{1mm}
\includegraphics[width=\columnwidth,keepaspectratio]{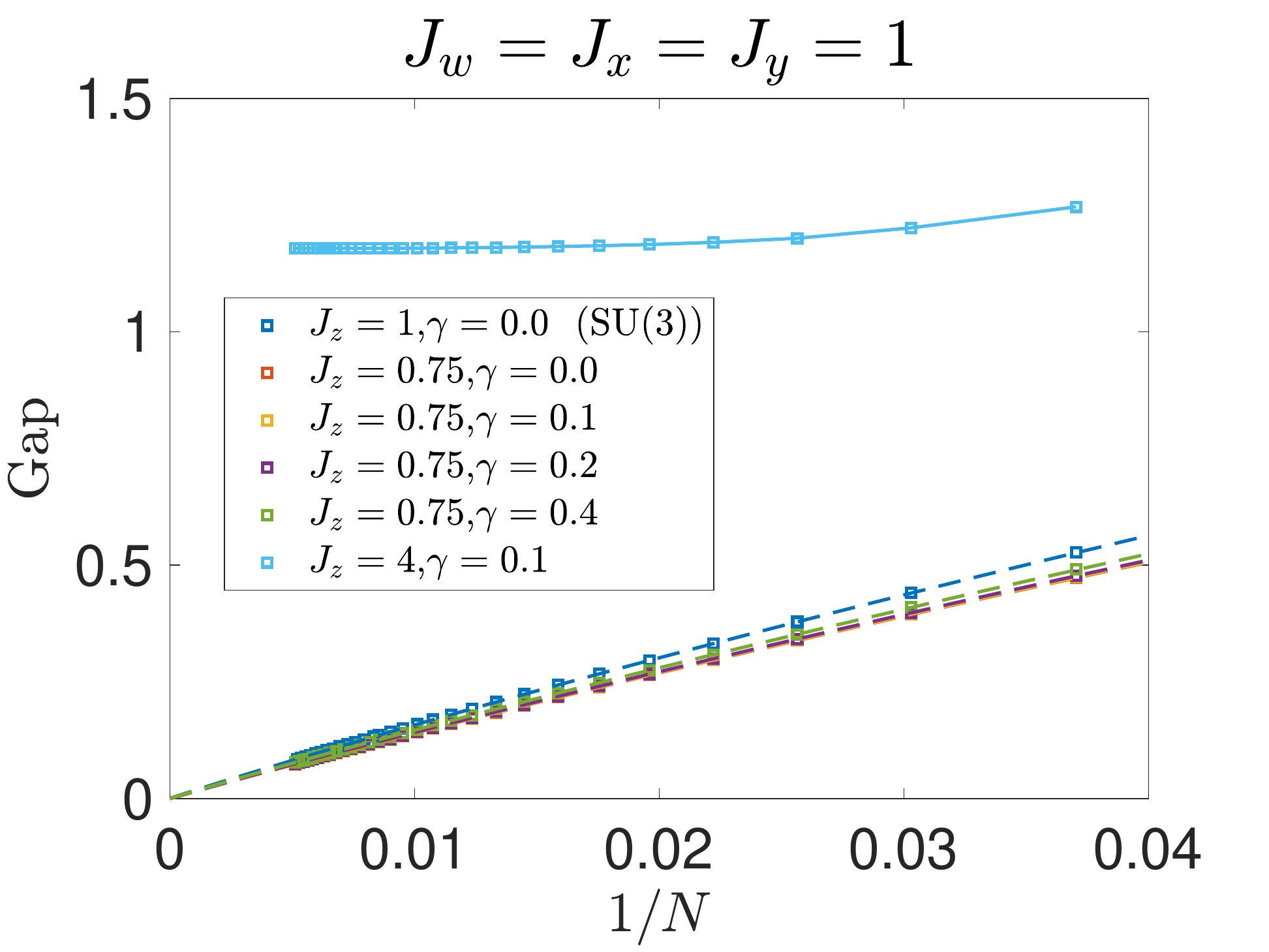} 
\caption{Finite size scaling of the energy difference between $S_z=0$ and $S_z =1$ sectors in the $-|J|<J_z<|J|$ gapless regime at $J=J_w=J_x=J_y=1$. We use open boundary condition and consider systems sizes up to $N=195$. One data set in  $|J|<J_z$ regime is included for comparison purposes.}
\label{gapgl1}
\end{figure}

As expected, we find that for $\gamma<0$ the system is in a gapped phase with a period $3$ anti-ferromagnetic ground state,
\begin{align}
\sum_j e^{2\pi i j/3} \langle Z_j \rangle   \neq 0 ~~\text{and}~~ \sum_j e^{-2\pi i j/3} \langle Z_j \rangle   = 0.
\end{align}
For $\gamma>0$ the system is in a different gapped anti-ferromagnetic state,
 \begin{align}
\sum_j e^{2\pi i j/3} \langle Z_j \rangle   = 0 ~~\text{and}~~ \sum_j e^{-2\pi i j/3} \langle Z_j \rangle   \neq 0.
 \end{align}
This further corroborates the fact that the field theory is in the $g > 1$ regime; otherwise this would be an irrelevant perturbation and should not gap the system for small values of $\gamma$. In Sec. \ref{g<1}, we show that in the regime that we expect $g < 1$, adding the $\gamma \neq 0$ term indeed does not gap the system, as expected (see Fig.~\ref{gapgl1}).
 
The two phases associated with $\gamma < 0$ and $\gamma > 0$ are symmetric with respect to two different symmetry subgroups $X T_x$ and $X^\dagger T_x$. Sample numerical results for the order parameters are shown in Fig.~\ref{fig4}. Within this picture, the inversion-symmetric $\gamma=0$ point is the critical point separating these two gapped phases. To confirm that this phase transition is not first order, we have also calculated the ground state energy as a function of $\gamma$. Results of this calculation are shown in Fig.\ref{fig5}, showing clear sign of a continuous phase transition. 

 To check consistency, we add a second nearest neighbor term and study the same phase transition with a nonzero $J_{2z}$. Again we find that for $\gamma<0$ and $\gamma>0$ the system is in different gapped phases ($g_1g_2g_3$ and $g_3g_2g_1$). However, the phase transition at the $\gamma=0$ point is now first order with a sixfold ground state degeneracy. In Figs.~\ref{g3} and \ref{g4} we have plotted the ground state energy as a function of $\gamma$ for two different values of $J_{2z}\neq 0$, showing clear signs of a first order phase transition.

We further remark that this phase transition is also present in a $\mathbb{Z}_3\times\mathbb{Z}_3$ symmetric integrable deformation of the $SU(3)$ spin chain,
\begin{align}
  \label{etaDeformation}
H=&\sum_i \Big( W_i W^\dagger_{i+1}+ X_i X^\dagger_{i+1} +  Y_i Y^\dagger_{i+1} \\ \newline \nonumber
&+(\cosh (\eta)+\frac{i}{\sqrt{3}} \sinh(\eta) ) Z_i Z^\dagger_{i+1}  \Big)+h.c~.
\end{align}
This spin chain has been exactly solved using Bethe-ansatz methods\cite{BABELON1,BABELON2,SCHULTZ,VEGA}. In agreement with our results, this system is gapped for $\eta>0$ and $\eta<0$ corresponding to the two gapped phases discussed above. The gap closes at the $PSU(3)$ symmetric critical point separating the two phases $\eta=0$. The continuous phase transitions discussed in this section, which occur in the regime $J_z > | J|$ at $\gamma = 0$ and away from the $PSU(3)$ symmetric point, are continuously connected to this $\eta = 0$ phase transition.

\subsubsection{Frustration and constrained Hamiltonians}

To help in understanding the effect of the frustration associated with $J_z$, we can consider the limit $J_z/|J| \rightarrow \infty$. In this limit, the $J_z$ term acts as a projection into its ground state subspace, forbidding neighboring spins from pointing in the same direction, $g_i\neq g_{i+1}$. Note that despite being exponentially large, this subspace has no local tensor product structure. This situation is somewhat similar to the Rydberg blockade effect in quantum simulators made of cold Rydberg atoms where two neighboring sites cannot both be excited\cite{rhydberg1,rhydberg3,rhydberg6}.

In the $J_w=J_x=J_y=J$ and $J_z/ |J| \rightarrow \infty$ limit, we can write an effective Hamiltonian,
 \begin{align}\label{schm}
 H_e= \mathcal{P}_0 \sum_i P_{i,i+1} \mathcal{P}_0,
 \end{align}
 where $\mathcal{P}_0$ is projection operator into the ground state subspace of the $J_z$ term, and $P_{i,i+1}$ is the permutation operator that swaps the state at $i$ and $i +1$. The quantum torus chain, Eq.\eqref{qtc}, is also described by Eq. \ref{schm} in the $\theta$ close to zero regime\cite{ryu1}. From the numerical results of the quantum torus chain, we can thus conclude that the projected Hamiltonian, Eq. \ref{schm}, also appears to be described by a $c = 2$ CFT. In other words, the nearest-neighbor spin chain with parameters $J_w=J_x=J_y=J$ remains gapless in the limit $J_z/ |J| \rightarrow \infty$. 
 
We remark that a similar effect has been found in constrained XXZ spin chains. That is, XXZ spin chains with a hardcore constraint that forbids two spin ups in two neighboring sites. This problem has been exactly solved using Bethe-ansatz methods in Ref.~\onlinecite{Alcaraz}. There, it was found that the system remains gapless and that the effect of the constraint is to simply renormalize the Luttinger parameter and the Fermi velocity. Using a similar idea, Ref.~\onlinecite{pollmann} was able to identify microscopic models without a $U(1)$ symmetry where
a $c=1$ Luttinger liquid phase with an \textit{emergent} $U(1)$ charge is stabilized. 

Therefore, since the $J_z$ term has an exponentially large degeneracy by itself, its effect on the low energy physics is not clear. Nevertheless the numerical results of the spin chain for finite values of $J$, together with the extreme limiting case $J_z/|J| \rightarrow \infty$ of Eq.\eqref{schm}, suggest that the system indeed remains gapless and is adiabatically connected to the $PSU(3)$ invariant point, which is described by the $SU(3)_1$ WZW CFT.

Finally we remark that for large enough values of $g$, other cosine operators, for example,
\begin{align}
\cos (3\phi^1)+\cos (3\phi^2)+\cos (3\phi^1+3\phi^2),
\end{align}
with scaling dimension $6/g$, will also become relevant, although still less relevant than $V_1$. Nevertheless, the arguments above indicate that the spin chain does remain gapless in the $J_z/|J| \rightarrow \infty$ limit. 

\subsection{Stability of the critical phase for $1/3 < g<1$}
\label{g<1}
\begin{figure*}[t]
    \centering
    \begin{subfigure}[t]{0.5\textwidth}
        \centering
        \vspace{0mm}
\includegraphics[width=\columnwidth,keepaspectratio]{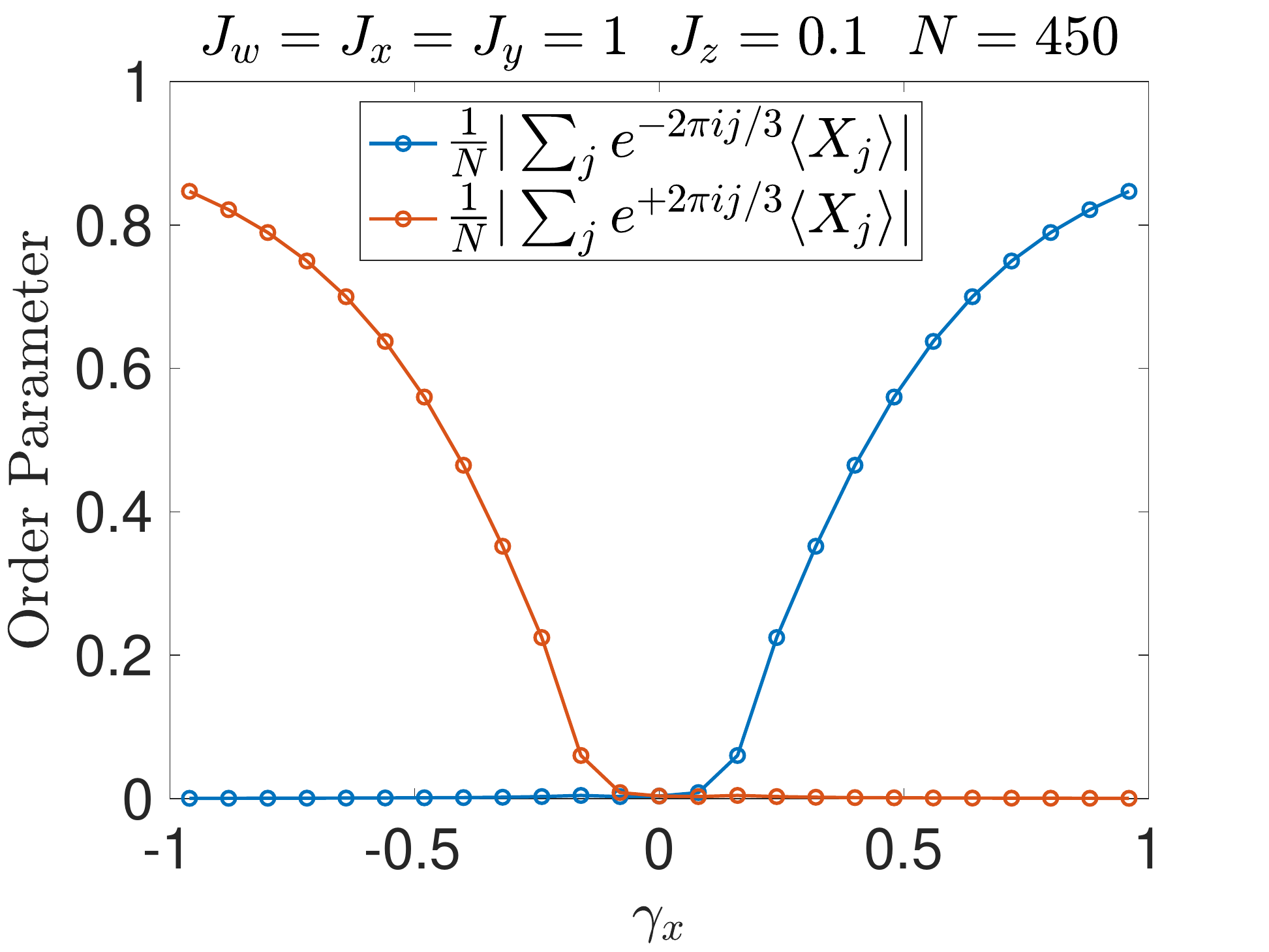} 
\caption{}
\label{odx}
    \end{subfigure}%
    ~ 
    \begin{subfigure}[t]{0.5\textwidth}
        \centering
       \vspace{0mm}
\includegraphics[width=\columnwidth,keepaspectratio]{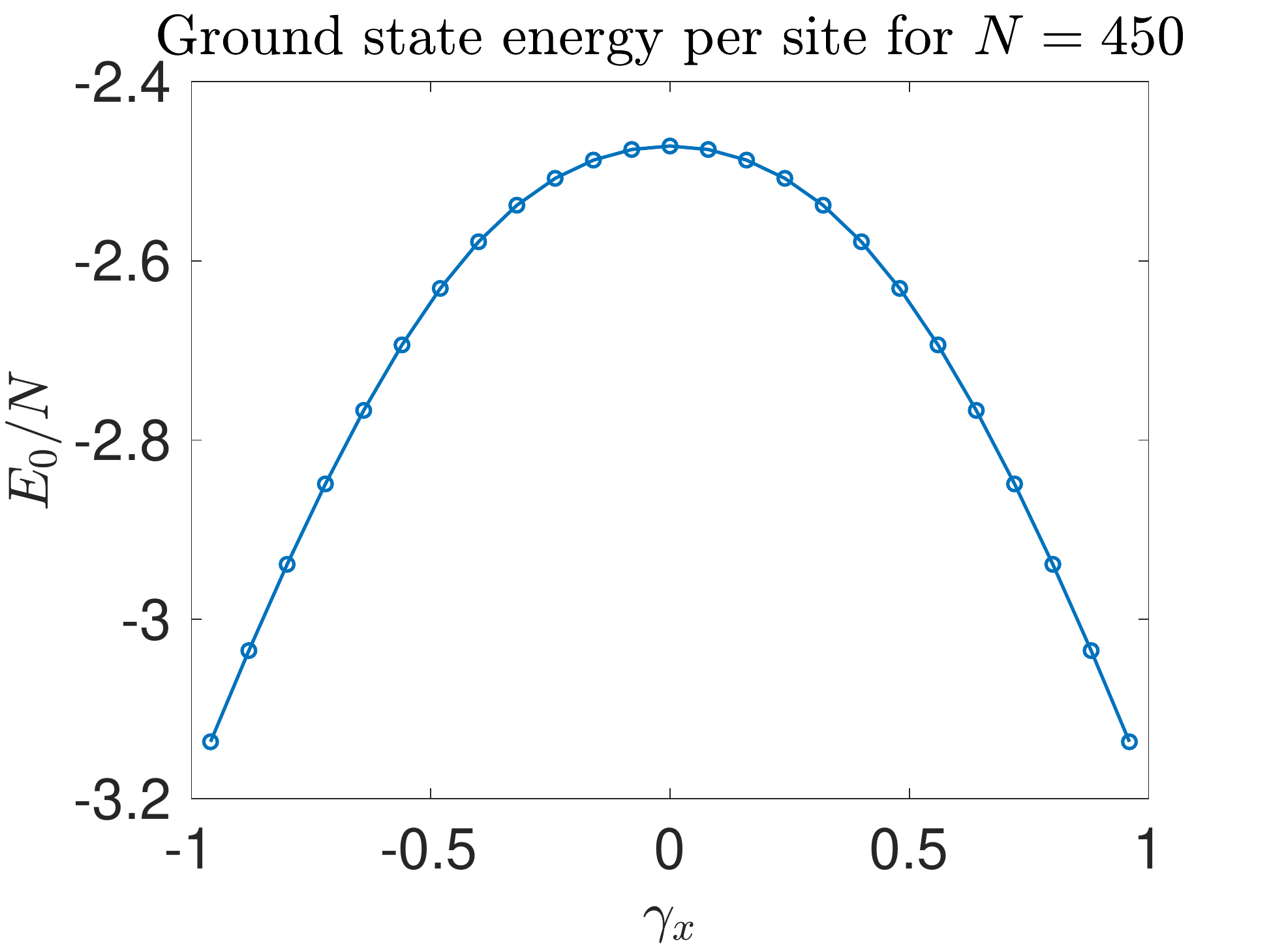} 
\caption{}
\label{egx}
    \end{subfigure}
    \caption{Order parameter and ground state energy as function of $\gamma_x$ (defined in Eq.\eqref{numh12}). Here we have set $J_w=J_x=J_y=1$ and $J_z=0.1$. The results are consistent with a continuous phase transition, with a gapped ordered phase developing for small non-zero $\gamma_x$. We have verified that the size of the region near $\gamma_x = 0$ where both order parameters are approximately zero itself decreases as the system size is increased, which helps provide evidence that indeed a small $\gamma_x$ does lead the system to spontaneously order and develop a gap.}
\end{figure*}

Here we first present evidence that the regime $g<1$ can be accessed by setting $-J < J_z< J$, with $J_x = J_y = J_w = J$. As discussed in the last section, this is expected from the bosonization of the $Z_i^\dagger Z_{i+1} + h.c.$ term near the $PSU(3)$ symmetric point. Subsequently, we provide evidence that the gapless nature of the spin chain continues to be stable as the $U(1)^2$ global symmetry is broken by further perturbing $J_x$ away from $J_y = J_w$. 

\subsubsection{Irrelevant $P$ and $\mathcal{C}$ symmetry breaking term from imaginary $J_z$}

As discussed in the previous section, we expect that $\tilde{V}_1$ would be generated by an imaginary $J_z$ term ($\gamma \neq 0$ in Eq. \ref{numh}), which breaks inversion, and charge-conjugation symmetry. Since $\tilde{V}_1$ has scaling dimension $2/g$, it is relevant for $g > 1$ but irrelevant for $g < 1$. Therefore, in the $g < 1$ regime, we expect that the system would remain gapless in the presence of a small $\gamma \neq 0$ term. We have indeed numerically confirmed that adding the $\gamma \neq 0$ term (Eq. \ref{numh}) indeed does not gap the system. Specifically, we observed that the finite-size scaling of the energy gap is consistent with a vanishing gap in the thermodynamic limit. Fig.~\ref{gapgl1} shows the energy difference between $S_z=0$ and $S_z =1$ sectors in this regime. We have used a fit of the form $\Delta=\frac{a}{N}+\frac{b}{N\log(N)}+c$, where the logarithmic corrections are induced by marginal operators. 

\subsubsection{Relevant $P$, $\Theta$, and $\mathcal{C}$ symmetry breaking term from imaginary $J_x$}

Let us now consider adding a symmetry-breaking imaginary part to the $J_x$ term,
\begin{align}
 i \sum_i (X_i X_{i+1}^\dagger-X_i^\dagger X_{i+1}).
\end{align}
This term is odd under time reversal, inversion and charge conjugation symmetries. The ground state of this term favors two different period-$3$ anti-ferromagnetic patterns depending on the sign of the term above, and is thus not frustrated. Therefore, we expect this term to destabilize the critical phase.

Based on symmetry properties, this term can generate the operator
\begin{align}
 \tilde{V}_2 = &\cos (\theta^1+\theta^2)+\cos (2\theta^2-\theta^1)+\cos (2\theta^1-\theta^2)- \\ \newline \nonumber
 &\cos (\phi^1-\phi^2-\theta^1-\theta^2)-\cos (\phi^1+2\phi^2+2\theta^1-\theta^2) \\ \newline \nonumber
         &-\cos (2\phi^1+\phi^2+\theta^1-2\theta^2),
\end{align}
which is a relevant perturbation for $1/3 < g < 1$ in the effective field theory.

Therefore, we expect that if the regime $-J < J_z< J$, with $J_x = J_y = J_w = J$ is described by the $1/3< g < 1$ regime of the Luttinger liquid theory, then a small imaginary part to $J_x$ should gap the system. To test our hypothesis numerically, we have simulated the spin chain with Hamiltonian 
\begin{align}\label{numh12}
H=\sum_i \Big( & J_w W_i W^\dagger_{i+1} +  J_y Y_i Y^\dagger_{i+1} + J_z Z_i Z_{i+1}^\dagger \\ \newline \nonumber
&+J_x(1+i \gamma_x ) X_i X^\dagger_{i+1}  \Big)+h.c~.
\end{align}
As expected, we find that for $\gamma_x\neq 0$ the system is in one of two gapped phases with a period $3$ anti-ferromagnetic ground state depending on the sign of $\gamma_x$. In Figs.~\ref{odx} and \ref{egx} we have plotted the order parameters and the ground state energy as a function of $\gamma_x$. This also shows clear signs of a continuous phase transition, which further indicates that the system is gapless when $\gamma_x = 0$. 

\subsubsection{Explicitly breaking $U(1)^2$ and persistence of gapless phase}

An obvious way to guarantee the stability of the critical phase in the $g <1$ regime is to enforce the microscopic $U(1)^2$ symmetry. In this case, $V_2$ and $V_3$ are forbidden and the entire $g<1$ phase is obviously stable. However, the system does not host \it emergent \rm charges, as the $U(1)^2$ charges are microscopically conserved. 

We would also like to understand the fate of the critical phase \it without \rm the microscopic $U(1)^2$ symmetry. Thus we consider perturbing the model with $J_w=J_x=J_y=J$ and $-|J|<J_z<|J|$ by tuning $J_x$ away from $ J_w = J_y = J$. Based on symmetry arguments, we observe that the $W_i W_{i+1}^\dagger+Y_i Y_{i+1}^\dagger+h.c$ and $X_i X_{i+1}^\dagger+h.c.$ operators, each of which individually break the microscopic $U(1)^2$ symmetry, can generate $V_2$. Therefore, naively the field theory would suggest that tuning $J_x$ away from $J_w = J_y = J$ in the regime $J_z < |J|$ would induce the relevant $V_2$ perturbation and destabilize the gapless phase.

Next, we note that the $X_i X_{i+1}^\dagger+h.c.$ term is frustrated, in that it has an exponentially large ground state degeneracy. Thus, by analogy with the analysis of the preceding section, we actually expect that tuning $J_x$ away from $J_w = J_y = J$ will also not destabilize the critical phase. Similarly, the $W_i W_{i+1}^\dagger+Y_i Y_{i+1}^\dagger+h.c$ term is itself gapless as it corresponds to the quantum torus chain (Eq.\eqref{qtc}) at $\theta=\pi/4$. We thus expect that the system remains gapless as $J_x$ and $J_w = J_y$ are tuned away from each other; this is numerically supported by the gapless nature of the system along the one-parameter flow described by Eq.\ref{Hlambda} and in the quantum torus chain (Eq.\eqref{qtc}). Specifically, the numerical study of Eq.\ref{Hlambda} shows that if we start with the $PSU(3)$ symmetric point and tune $J_x, J_z$ continuously to zero to obtain the quantum torus chain, the system stays gapless.

We do not discuss the $0<g<\frac{1}{3}$ regime where the most relevant operator is given by $V_3$ in Eq.\eqref{pert} and which is allowed in principle if the $U(1)^2$ symmetry is broken. We have not studied whether the microscopic model accesses this regime of the field theory.

\section{Nearest neighbor Hamiltonian with $\mathbb{Z}_n\times\mathbb{Z}_n$ symmetry}
\label{ZnSec}

In this section, we generalize the discussion above to the case of  $\mathbb{Z}_n\times\mathbb{Z}_n$ symmetry for arbitrary $n>2$. We have not studied these systems numerically for $n>3$. Nevertheless, as we argue below, it is consistent and natural to expect that for $n$ odd the phase diagrams of these models will also have stable gapless phases described by $n-1$ component Luttinger liquids. For even $n  > 2$, our considerations further suggest that nearest neighbor $\mathbb{S}_n \times \mathbb{Z}_n$ symmetric spin chains may also harbor gapless phases for large regions of their phase diagrams. 

Similar to the construction of Eq.\eqref{h3}, we can write the most general nearest neighbor $\mathbb{Z}_n\times \mathbb{Z}_n$ symmetric Hamiltonian in terms of clock variables. At the maximally symmetric (antiferromagnetic) point, this spin chain will have a $PSU(n) = SU(n)/\mathbb{Z}_n$ on-site symmetry. At this point the low energy description is given by the $SU(n)_1$ WZW CFT\cite{wzw0,wzw1,wzw2}. We can then modify the parameters around the $SU(n)$ symmetric point to break the $PSU(n)$ symmetry down to $\mathbb{Z}_n\times\mathbb{Z}_n$. We emphasize that these systems still obey the same LSM restrictions. Additionally, similar to the discussion of $\mathbb{Z}_3\times \mathbb{Z}_3$, there exists $\mathbb{Z}_n\times \mathbb{Z}_n$ symmetry-preserving marginal deformations of the $SU(n)_1$ WZW CFT, which give rise to $n-1$-component Luttinger liquids. These Luttinger liquids have the same mixed LSM anomalies as the microscopic lattice models and are thus also natural candidate theories to describe the gapless phases of these systems.
 
Below we show how to explicitly see the LSM anomaly in these $n-1$ component Luttinger liquids. Similar to the previous cases, the form of the fixed point action is given by the sigma model action of Eq.\eqref{string2}.  For the $SU(n)$ invariant point we can choose\cite{giveon1},
 \begin{align}
 G^{SU(n)}_{ij}=\frac{1}{4}~~\text{for}~~i\neq j,~~~\text{and}~~~ &G^{SU(n)}_{ii}=\frac{1}{2} \\ \newline \nonumber
 B^{SU(n)}_{ij}=\frac{1}{4}~~\text{for}~~i> j,~~~\text{and}~~~ &B^{SU(n)}_{ij}=\frac{-1}{4} ~~\text{for}~~i< j.
 \end{align}
 We also define the Luttinger variables $\theta^i,\phi^i$ as in Eq.\eqref{lutt}.
 
We can again use abelian bosonization at the $SU(n)$ invariant point to find the symmetry action on the fields in the effective long wavelength field theory. This calculation is analogous to the $\mathbb{Z}_3\times \mathbb{Z}_3$ case presented in the Appendix.~\ref{spm}. Here we just quote the results.

The action of the symmetry associated with $X$ is
\begin{align}
  X:~~&\phi^i \rightarrow M_{ij} \phi^j
\nonumber \\
  &\theta^i \rightarrow (M^T)^{-1}_{ij} \theta^j   ,
\end{align}
for
\begin{align}
M=\left(\begin{array}{cccc} -1 & -1 & -1 & \cdots \\ +1 & 0 & 0 & \cdots \\  0 & +1 & 0 & \cdots  \\ 0 & 0 & +1 & \cdots \end{array}\right),
\end{align}
Again we can introduce an extra field $\varphi^n_{L/R}$, such that $\sum_{i=1}^n \varphi^i_{L/R} = 0$, in which case $X$ acts as a $\mathbb{Z}_n$ cyclic permutation of $\varphi$ fields. 

The action of the symmetry associated with $Z$ is
\begin{align}
  Z:~~&\theta^\alpha \rightarrow \theta^\alpha +  2\pi \alpha/n
  \nonumber \\
  &\phi^i \rightarrow  \phi^i .\end{align}
The action of translation symmetry by one site, $T_x$, is given by,
\begin{align}
  T_x: ~~ &\theta^\alpha\rightarrow \theta^\alpha+ 2\pi \alpha/n
  \nonumber \\
  &\phi^\alpha\rightarrow \phi^\alpha+2\pi/n.
\end{align} 
Inversion $P$ acts as
 \begin{align}
   P: ~~ &\theta^i\rightarrow \theta^i - 4 B^{SU(n)}_{ij} \phi^j
   \nonumber \\
   &\phi^i\rightarrow -\phi^i.
\end{align}
Finally, time-reversal (complex conjugation in $Z$ basis) acts as
 \begin{align}
   \Theta:~~ &\theta^i\rightarrow -\theta^i + 4 B^{SU(n)}_{ij} \phi^j
\nonumber \\
   &\phi^i\rightarrow \phi^i .
\end{align}

Similar to the $\mathbb{Z}_3\times \mathbb{Z}_3$ case, inversion and time-reversal fix the $B$ matrix such that $4B$ is an integer matrix. However, the $G$ matrix can vary as long as $M^T G M =G$.

To see the the LSM anomaly, we insert a unit of $T_x$ flux. This changes the boundary conditions such that 
 \begin{align}\label{phit2}
&\phi^\alpha(x+2\pi)=\phi^\alpha(x)+2\pi m^\alpha +  2\pi/n\\ \newline \nonumber
&\theta^\alpha(x+2\pi)=\theta^\alpha(x)+2\pi n^\alpha+  2\pi \alpha/n.
\end{align}
This in turn can be absorbed into a shift of $n^i$ and $m^i$,
\begin{align}
m^\alpha\rightarrow m^\alpha + \frac{1}{n} ~~ \text{and} ~~  n^\alpha\rightarrow n^\alpha + \frac{\alpha}{n}.
\end{align}
 As shown above, inserting a flux of translation symmetry through the system then fractionalizes the allowed values of $U(1)^n$ charges ($n^i,m^i$). Since the total charge cannot be zero, in this case the ground state is at least $n$ fold degenerate. The ground states are generated by the transformations
\begin{align}\label{xzn}
 \left(\begin{array}{c} n \\ m \end{array}\right) \rightarrow \left(\begin{array}{cc} (M^T)^{-1} & 0 \\ 0 & M\end{array}\right)  \left(\begin{array}{c} n \\ m\end{array}\right).
\end{align}

 We now proceed to discussing the stability of the critical phase. The generic perturbations that could potentially gap the bosonic fields are the vertex operators 
\begin{align}
\mathcal{O}_{mn} = \cos (m^i\theta^i+n^i\phi^i),
\end{align}
with scaling dimension given by Eq.\eqref{0m}. Enumerating all allowed operators is more complicated in the $SU(n)$ case. However, at least in the particular case of $G=g G^{SU(n)}$ with $g>1$, the most relevant symmetry allowed terms are $n-2$ distinct operators of the form,
 \begin{align}
V_j = &\cos (\phi^1-\phi^{1+j})+\text{terms related by $X$ symmetry} \\ \newline \nonumber
&~~~\text{for } j=1, \cdots, n-2.
\end{align}
 These operators are related by the permutation group $\mathbb{S}_n$. At the $SU(n)$ symmetric point $g=1$, all such operators are marginal. As we increase $g$ these operators become relevant. Note that as long as the $G$ matrix is proportional to $G^{SU(n)}$, we have an enlarged symmetry group $\mathbb{S}_n\times \mathbb{Z}_n$. In this case, only the sum of all the cosine operators above can appear; due to incompatibility of arguments (with a positive sign), all of the cosines  cannot be simultaneously minimized. It is easy to come up with a microscopic term with the same symmetries that has an exponentially large $d_g=n (n-1)^{N-1}$ ground state degeneracy ($N$ is the number of sites). The physics here is analogous to the $\mathbb{Z}_3\times \mathbb{Z}_3$ case with $J_z\neq1$. We thus expect stable critical phases with emergent charges to be also present in this case. 
 

 We can then break the $\mathbb{S}_n\times \mathbb{Z}_n$ symmetry all the way down to $\mathbb{Z}_n\times \mathbb{Z}_n$ by deforming the $G$ matrix so that it is not proportional to $G^{SU(n)}$ (this is possible for $n>3$). In this case, the operators above do not have the same scaling dimension. Without any loss of generality, assume that the most relevant operator is
  \begin{align}
\sum_{i=1}^{n}\cos (\phi^i-\phi^{i+1}),
\end{align}
where the auxiliary field $\phi^n=-\sum_{i=1}^{n-1} \phi^i$. In this case, if $n$ is odd, all of the cosines cannot be simultaneously minimized. We can write a microscopic term with this symmetry $Z_iZ_{i+1}^\dagger+h.c.$ that has an exponentially large $d_g=n 2^{N-1}$ ground state degeneracy ($N$ is the number of sites). We expect stable critical phases to be also present in this case. However, if $n$ is even, the arguments are compatible, and all the cosine term can be simultaneously minimized, e.g. $\phi_{2i}=0$ and $\phi_{2i}=\pi$, giving a rise to a conventional gapped antiferromagnetic phase. The associated microscopic term is also not frustrated in this case.
 
 Bases on this argument, we believe that extended critical phases with emergent charges should be present in generic $\mathbb{Z}_{2n+1}\times \mathbb{Z}_{2n+1}$ spin chains with time-reversal and inversion symmetries. 

 \section{Discussion}\label{s5}
 
In this paper, we have argued that a wide class of translationally invariant $\mathbb{Z}_3 \times \mathbb{Z}_3$ symmetric spin chains are described at long wavelengths by a two-component Luttinger liquid. Importantly, the Luttinger liquid, with the appropriate symmetry actions, reproduces the mixed anomaly associated with the generalized LSM constraint of this system. With time-reversal, inversion or charge-conjugation symmetry, the two-component Luttinger liquid theory has a single continuous free parameter. Remarkably, despite the apparent presence of symmetry-allowed relevant operators in the field theory away from the $SU(3)$ invariant point, the system appears to remain gapless for a large range of microscopic parameters. These systems therefore host, at least to an excellent approximation, \it emergent \rm $U(1)^2$ conserved charges, as the microscopic model only has discrete global symmetries. Remarkably, these emergent charges seem to continue to exist throughout a large portion of the phase diagram. We discussed the importance of microscopic frustration in stabilizing the gaplessness of the model. 

These results are rather surprising, and suggest a number of possible scenarios that at present we cannot distinguish. First, it is in principle possible that the gapless (or near gapless) nature of the system for an essentially infinite range of dimensionless microscopic parameters is accidental, in the sense that the RG flows happen to not generate the relevant operators, or do so with extremely small amplitudes. However such a scenario, while conceptually consistent, is in tension with the notion of naturalness in effective field theory. On the other hand, it is conceivable that the nearest neighbor microscopic Hamiltonians we consider have additional conserved quantities, such as due to a previously unnoticed integrability or partial integrability that is present in the lattice model even away from the $SU(3)$ point. In this case, the relevant operators may be forbidden due to the associated hidden symmetries of the model, which might generically be broken by adding next neighbor terms. 

We have also shown that the mixed anomaly associated with the generalized LSM constraint in translationally invariant $\mathbb{Z}_n \times \mathbb{Z}_n$ spin chains is also satisfied for $c = n-1$ multi-component Luttinger liquids with appropriate symmetry actions. We provided a series of arguments for why stable critical phases with emergent charges may be expected to exist in: (a) $\mathbb{Z}_n \times \mathbb{Z}_n$ symmetric spin chains with $n$ odd, and (b) $\mathbb{S}_n \times \mathbb{Z}_n$ symmetric spin chains with all $n > 2$. It would be interesting to study these systems (with $n>3$) in more detail in future work.

Inspection of the $c = n-1$ multi-component Luttinger liquids theories shows that it does not appear to be possible to gap any of the $n-1$ modes while preserving the $\mathbb{Z}_n \times \mathbb{Z}_n$ and translational symmetries. This raises an interesting question of what is the minimum central charge $c_{min}$ of field theories that can possess such a mixed anomaly and, in particular, whether there is a universal lower bound to $c_{min}$ that grows with $n$. The results of this paper are consistent with $c_{min} \geq n-1$ and thus raise the question of whether there is any theory with $c < n-1$ that has the appropriate mixed anomaly. We note that a bound on $c_{min}$ for such mixed LSM anomalies is closely related to a bound recently conjectured in Ref.~\onlinecite{cbound1} for the central charge of CFTs that appear at phase transitions between gapped SPT states. We also note that Ref. \onlinecite{lin2019} recently established bounds on the number of charged degrees of freedom from anomalies in (3+1)D superconformal field theories. 

Another interesting feature of the theories studied here is that they offer the possibility of having distinct CFTs with the same central charge and symmetries that are described with different topological $B$ matrices. Interestingly all of these theories (with different $B$'s) also reproduce the same LSM anomaly. It would be interesting to understand whether both of these phases can exist in the same system and if so, to study their domain walls. 

\section*{Appendix: Abelian Bosonization to Derive Symmetry Actions}\label{spm}

Here we review the derivation of the Abelian bosonization for the $SU(3)$ spin chain, which can be used to derive the symmetry actions on the scalar fields of the $c=2$ Luttinger liquid. We then use the same symmetry action on the fields throughout the moduli space of the $\mathbb{Z}_3 \times \mathbb{Z}_3 \times \mathbb{Z}_{\text{trans}}$ symmetric models. All of the calculations of this section are straightforwardly generalizable to the case of $SU(n)$.

Interestingly, the bosonization, even at the $PSU(3)$ symmetric point, has an inherent ambiguity that is related to the question of whether $b = 0$ in the expansion of Eq. \ref{fttt}.

Following Refs.~\onlinecite{affleck,affleck2,affleck3}, we start with the Hamiltonian for the SU(3) Hubbard model,
\begin{align}
H= -t\sum_{\alpha=1}^3\sum_i [\psi^{\alpha\dagger}_i \psi^{\alpha}_{i+1}+h.c.]+U \sum_i [\sum_{\alpha=1}^3\psi^{\alpha\dagger}_i \psi^{\alpha}_{i}-1]^2.
\end{align}
Here $\psi^\alpha$ are the usual fermionic annihilation operators transforming in the fundamental representation of $SU(3)$. In the large coupling limit $U/t\gg 1$ where the band is $1/3$ filled, this Hamiltonian maps onto the $PSU(3)$ symmetric spin chain described in Eq.\eqref{su3}. In the non-interacting limit $U = 0$, we can write a continuum expansion for the fermion fields around the Fermi wave vector $k_F = \pm\pi/3$ as
\begin{align}\label{trans}
\psi^\alpha_j\approx [e^{-i\pi j/3} \psi^\alpha_{L}(x_j)+e^{+i\pi j/3} \psi^\alpha_{R}(x_j)],
\end{align}
where $x=j\mathfrak{a}$ and $\mathfrak{a}$ is the lattice spacing. From here on, we work in dimensionless length units where $\mathfrak{a}=1$.
In this limit, the low energy field theory is the $U(3)_1$ WZW model. In the Abelian bosonization approach, we can bosonize the fermion fields as
\begin{align}
  \psi^\alpha_{L} (x) &\propto :e^{-i \tilde{\varphi}^\alpha_L (x)}: ,
\nonumber \\
  \psi^\alpha_{R} (x) &\propto :e^{+i \tilde{\varphi}^\alpha_R (x)}:.
\end{align}

We assume that the limit $U/t \rightarrow \infty$ can be described by constraining the occupation number at each site:
\begin{align}\label{cnst2}
\sum_\alpha\big(&\psi^{\dagger\alpha}_L (x_j) \psi^{\alpha}_L (x_j)+\psi^{\dagger\alpha}_R (x_j) \psi^{\alpha}_R (x_j) + \\ \newline \nonumber
& (e^{+2 \pi i j/3} \psi^{\dagger\alpha}_L (x_j) \psi^{\alpha}_R (x_j) + h.c)\big)=0.
\end{align}
In the bosonic language, this gives the $3$ constraints,
\begin{align}\label{cnst1}
  \sum_\alpha\partial_z \tilde{\varphi}^\alpha_L&=0
    \nonumber \\
  \sum_\alpha\partial_{\bar{z}} \tilde{\varphi}^\alpha_R&=0
  \nonumber \\
\sum_\alpha \sin(2\pi j/3+\tilde{\varphi}^\alpha)&=0,
\end{align}
where $\tilde{\varphi}^\alpha=\tilde{\varphi}^\alpha_L+\tilde{\varphi}^\alpha_R$. The first $2$ constraints can be satisfied by setting $\tilde{\varphi}_L^1+\tilde{\varphi}_L^2+\tilde{\varphi}_L^3=0$ (and similarly for $\tilde{\varphi}_R$). This constraint gaps out the $U(1)_1$ total charge mode, leaving behind the $SU(3)_1$ WZW model describing the spin degrees of freedom. However note that, using this relation alone, the large $U$ constraint is only imposed at a mean-field level, as only the non-oscillating part of Eq.\eqref{cnst2} vanishes. 

For the $SU(2)$ case, the third constraint is also automatically satisfied. However, it is not clear how to fully incorporate the effects of the third constraint in the bosonization scheme for $n>2$. Following Ref.~\onlinecite{affleck3} we ignore this constraint from here on.

The conserved $SU(3)$ currents are given by\cite{affleck},
\begin{align}\label{afcur}
  J_{L}^a &\propto \sum_{\alpha,\beta=1}^{3} \psi^{\alpha\dagger}_L \lambda^a_{\alpha\beta} \psi^{\beta}_L
\nonumber \\
    J_{R}^a &\propto \sum_{\alpha,\beta=1}^{3} \psi^{\alpha\dagger}_R \lambda^a_{\alpha\beta} \psi^{\beta}_R,
\end{align}
where $\lambda^a$ are the Gell-Mann matrices.

The microscopic spin-chain operators $\lambda^a_j$  can be written as follows. First, we note that in the $SU(3)$ Hubbard model, the spin operators  $\lambda^a_j$
are written as
\begin{align}
\lambda^a_j = \psi^{\alpha\dagger}_j \lambda_{\alpha\beta}^a \psi^\beta_j. 
  \end{align}
When passing to the continuum, we then obtain
  \begin{align}\label{afcur1}
\lambda^a_j\propto J_{L}^a +   J_{R}^a + \sum_{\alpha,\beta=1}^{3} [e^{2\pi i j /3} \psi^{\dagger\alpha}_L  \lambda^a_{\alpha\beta} \psi^{\beta}_R +h.c.].
\end{align}

Using the fact that $X_j$ and $Z_j$ are linear superpositions of the $\lambda_j^a$, we can then obtain their expression in terms of the long wavelength bosonic fields.

 The fixed point part of the field theory Hamiltonian can be written in the Sugawara (current squared) form,
\begin{align}
H\propto\int dx[ (J_L^3)^2 + (J_L^8)^2 + L \rightarrow R]
\end{align}
Note that $\lambda^{3,8}$ generate the Cartan sub-algebra of $SU(3)$ and can be explicitly written down as
\begin{align}
&J_L^3\propto \partial_z (\tilde{\varphi}^1_L-\tilde{\varphi}^2_L) \\ \newline \nonumber
&J_L^8\propto\sqrt{3}\partial_z (\tilde{\varphi}^1_L+\tilde{\varphi}^2_L).
\end{align}

Matching the two equations above with the Hamiltonian in the main text Eq.\eqref{hlr} we can make the identification
\begin{align}
(\varphi^1_L, \varphi^2_L) = (\tilde{\varphi}^1_L , \tilde{\varphi}^2_L).
\end{align}

As stated above, using Eq.\eqref{afcur1}, we can obtain expressions for microscopic operators. For example,
\begin{align}\label{abc1}
&Z_j\sim \sum_{\alpha=1}^3  e^{2\pi i \alpha/3} \big(\frac{1}{2\pi}  \partial_x \varphi^\alpha + c \sin(2\pi j/3 +\varphi^\alpha) \big),
\end{align}
where $c$ is a non-universal short distance physics dependent number.

The symmetry actions can now be read off from their action on the microscopic operators (e.g. $Z_j$ and $X_j$). For example, Eq. \ref{abc1} shows that a translation by one site, $j \rightarrow j+1$, requires $\varphi^\alpha \rightarrow \varphi^\alpha + 2\pi/3$. The action on $\theta^\alpha$ can similarly be obtained by considering the expansion for $X_j$, with the result reported in the main text. Note that these symmetry actions are not necessarily unique in terms of $\varphi_L^i$ and $\varphi_R^i$, however, their action on physical (non-chiral) variables $\phi^i$ and $\theta^i$ is unique. 

As mentioned in the beginning of this section, we assume that the symmetry actions remain the same throughout the moduli space. This must be the case for the $X$, $Z$, $\mathcal{C}$, $\Theta$, and $P$ symmetries, as they are all finite order discrete symmetries, and the only continuous symmetry of the two-component Luttinger liquid theory (for generic $g$) corresponds to $U(1)_L^2 \times U(1)_R^2$. There is thus no possible way to continuously deform the action of these symmetries while respecting the fact that they have finite order.

Since translation symmetry is of infinite order, it is in principle possible that the action of translation symmetry $T_x$ changes continuously, such that $T_x$ does not act as $\mathbb{Z}_3$ away from the $PSU(3)$ point. However, we have explicit numerical evidence (see Fig.\ref{spec.pdf}) that the action of $T_x$ remains as $\mathbb{Z}_3$ even far away from the $PSU(3)$ point. Furthermore, even if one imagines that the action of translation symmetry is modified in some portion of the phase diagram, this cannot rule out the appearance of the operators $V_1$, $V_2$ and $V_3$ unless the action of translation symmetry is different for $\varphi^1$ and $\varphi^2$, in which case, the $X$ symmetry would be necessarily broken. However, again we have a significant amount of numerical evidence that the ground state is in fact $X$ invariant throughout the gapless phase.

\subsection{Ambiguity in the bosonization scheme}

Using Eq.\eqref{abc1} we can easily derive Eq.\eqref{fttt} of the main text, with $a=\frac{3}{(2\pi)^2}$ and $b=(\frac{c}{2})^2$ ($c$ defined in Eq.\eqref{abc1}). However, we remark that there is an alternative way to bosonize this term by writing it in terms of fermionic variables as
\begin{align}
  Z_j Z_{j+1}^\dagger&+  h.c.
                       \nonumber \\
                       =& -\sum_{\alpha\neq\beta} \psi^{\dagger\alpha} (x_j) \psi^{\alpha} (x_j)  \psi^{\dagger\beta} (x_{j+1}) \psi^{\beta} (x_{j+1}) \\ \newline \nonumber  
& +2 \sum_{\alpha} \psi^{\dagger\alpha} (x_j) \psi^{\alpha} (x_j)  \psi^{\dagger\alpha} (x_{j+1}) \psi^{\alpha} (x_{j+1}) \\ \newline \nonumber
=& -1 +3 \sum_{\alpha} \psi^{\dagger\alpha} (x_j) \psi^{\alpha} (x_j)  \psi^{\dagger\alpha} (x_{j+1}) \psi^{\alpha} (x_{j+1}).
\end{align}
In going from the second line to the third, we used the large $U$ Hubbard constraint $\sum_\alpha  \psi^{\dagger\alpha} (x_j) \psi^{\alpha} (x_j)=1$. The final expression can now be bosonized to give,
\begin{align}
\sum_j Z_j Z_{j+1}^\dagger+  Z_j^\dagger Z_{j+1} \sim \int dx \frac{3}{(2\pi)^2} G_{ij} \partial_x \varphi^i \partial_x \varphi^j  .
\end{align}
As compared with Eq.\eqref{fttt}, the above equation does not include the perturbation term $V_1$. This suggests that the term $\sum_j Z_j Z_{j+1}^\dagger+h.c$ term might in fact not generate $V_1$ in the continuum limit ($b = 0 $ in Eq. \ref{fttt}). This ambiguity in the bosonization can be traced back to the third constraint in Eq.\eqref{cnst1}, which was ignored at the mean-field level. We again emphasize that a similar ambiguity does not exist for the $SU(2)$ case as the additional constraint is automatically satisfied.

\section*{Acknowledgments}

We thank B. Swingle, J. Sau, I. Affleck, A. Rahmani, P. Fendeley, S. Ryu and L. Motrunich for discussions. MB is supported by NSF CAREER (DMR- 1753240), an Alfred P. Sloan Research Fellowship, and JQI-PFC-UMD. YA is supported by National Science Foundation NSF DMR1555135 and JQI-NSF-PFC. We acknowledge the University of Maryland supercomputing resources (http://hpcc.umd.edu) made available for conducting the research reported in this paper.

\bibliography{library}

\end{document}